\documentclass{article}

\usepackage{arxiv}

\usepackage[utf8]{inputenc} 
\usepackage[T1]{fontenc}    
\usepackage{hyperref}       
\usepackage{amsthm} 

\usepackage{url}            
\usepackage{booktabs}       
\usepackage{amsfonts}       
\usepackage{nicefrac}       
\usepackage{microtype}      
\usepackage{lipsum}
\usepackage{graphicx}
\graphicspath{ {./images/} }

\usepackage{mathtools}
\usepackage{amssymb}
\DeclareMathOperator{\Tr}{Tr}

\usepackage[capitalize,nameinlink,noabbrev]{cleveref} 

\crefname{theorem}{Theorem}{Theorems}
\crefname{definition}{Definition}{Definitions}
\crefname{proposition}{Proposition}{Propositions}
\crefname{lemma}{Lemma}{Lemmas}
\crefname{corollary}{Corollary}{Corollaries}

\makeatletter
\@addtoreset{equation}{section} 
\makeatother

\newcommand{\matr}[1]{\boldsymbol{\mathbf{#1}}}  
\newcommand{\Lapl}{\matr{L}}
\newcommand{\p}{\matr{p}}
\newcommand{\T}{\matr{T}}

\newcommand{\dkl}{D_{\operatorname{KL}}}

\title{Conditional Entropy of Heat Diffusion on Temporal Networks}

\author{
 Samuel Koovely \\
  Department of Mathematical Modeling and Machine Learning\\
  University of Zurich\\
  Zurich, Switzerland \\
  \texttt{samuel.koovely@math.uzh.ch} \\
   \And
 Alexandre Bovet \\
  Department of Mathematical Modeling and Machine Learning\\
  University of Zurich\\
  Zurich, Switzerland\\
  \texttt{alexandre.bovet@math.uzh.ch}
}

\begin{document}
\maketitle
\begin{abstract}
Diffusion-based information-theoretic approaches provide new theoretical and practical tools to study complex networks. 
So far, they have not been generalized to temporal networks. In this work, we show that common entropic measures based on modeling diffusion on graphs, such as the entropy rate and the spectral entropy, do not generalize straightforwardly to temporal networks due to the process's non-stationarity and temporal asymmetries.
Instead, we propose the conditional entropy of heat diffusion as an entropic measure for continuous-time temporal networks and study its properties. 
We show that this quantity is monotone in time, yielding an information-theoretic analog of the second law of thermodynamics for inhomogeneous diffusion on temporal networks. We provide an upper bound and suggest a lower bound on its evolution and explain how discrepancies from it arise due to asymmetric temporal paths.
We then introduce a local version of conditional entropy, designed to probe diffusion over finite temporal windows, and show that it provides an informative signal for change-point detection in continuous-time temporal networks.
We evaluate the proposed methodology on synthetic benchmarks, including comparative experiments with existing nonparametric baselines in the snapshot setting, and then apply it to a real-world temporal contact network. Finally, we show how to use detected change points to guide community detection on targeted sub-intervals, improving the quality and interpretability of the clustering results.
\end{abstract}


\section{Introduction} \label{sec:intro}
Complex systems are ubiquitous in nature, and complex networks provide a flexible framework to model many of them \cite{newman_networks_2018}. The interactions among the constituents of a network determine its structural organization and shape its macroscopic behavior \cite{newman_structure_2003}. A natural way to study these interactions is through diffusion processes on the network, governed by the graph Laplacian \cite[Ch. 2.4.1]{lambiotte_modularity_2021}. At the global level, the associated heat kernel \cite[Ch. 2.5.2]{lambiotte_modularity_2021} captures how information spreads across the system.
Heat diffusion on a graph can be viewed as a time-homogeneous Markov process \cite{masuda_random_2017}, and constitutes the graph analog of heat diffusion in a physical medium. Its evolution is tightly linked to the underlying network structure: the Laplacian encodes key structural information through its eigenvalues and eigenvectors \cite{villegas_laplacian_2022}. This connection has been exploited in several directions, for instance, in the Markov stability framework for community detection \cite{delvenne_stability_2010, schaub_markov_2012, lambiotte_random_2014}. It has also been studied from an information-theoretic viewpoint. Several entropy-based quantities \cite{yang_unified_2020} have been proposed for networks, including entropy production \cite{schnakenberg_network_1976,nartallo-kaluarachchi_broken_2024}, Markov chain entropy rate \cite{gomez-gardenes_entropy_2008}, and the spectral entropy \cite{de_domenico_spectral_2016,ghavasieh_statistical_2020,villegas_laplacian_2022,villegas_laplacian_2023,villegas_multi-scale_2025},  which adapts the notion of Von Neumann entropy to networks and has provided new tools for graph renormalization and clustering. 
In this formalism, the diffusion time serves as a resolution parameter, allowing one to study the network from small to global scales.
In a recent work \cite{koovely_evolution_2025}, we introduced the conditional entropy of diffusion dynamics on static networks, showing that this quantity does not decrease in time and therefore satisfies the second law of thermodynamics. Additionally, we derived exact formulas for its finite-time evolution on certain classes of regular graphs, showing that it can be interpreted as an observable that captures structural properties of the underlying graph, including degree heterogeneity, small-world structure, and density.

Many real-world networks evolve in time, and so do their structural properties. Temporal networks \cite{holme_temporal_2012, masuda_guide_2016} are designed to model this setting. 
Thermodynamic characterizations of time-evolving networks have been proposed based on quantum mechanics and statistical physics \cite{yeThermodynamicCharacterizationNetworks2015,yeThermodynamicsTimeEvolving2015,minelloThermodynamicCharacterizationTemporal2016}. They consider temporal networks as sequences of independent static snapshots and derive thermodynamic quantities at each time step. Here, we are interested in a more general description of temporal networks evolving in continuous time, the so-called link streams \cite{latapy_stream_2018}, and in modeling a diffusion process that evolves with the network, allowing us to derive entropic measures intrinsically linked to the network's evolution.
Heat diffusion generalizes naturally to temporal networks as a continuous-time inhomogeneous Markov process \cite{bovet_flow_2022}.  In this formalism, the diffusion rate is the dynamical resolution parameter.
However, we show that the Markov Chain entropy rate and the von Neumann-type spectral entropy do not admit straightforward extensions to this setting. This motivates the study of other diffusion-based information-theoretic quantities in the temporal setting. In this article, we extend the conditional entropy of heat diffusion to temporal networks and show that the monotonicity property underlying its second-law interpretation in the static case \cite{koovely_evolution_2025} still holds in the temporal case.

We then introduce a local version of conditional entropy and use it to identify change points, that is, times at which the network structure changes abruptly. From a signal-processing perspective, a temporal network may be viewed as a trajectory in an abstract non-Euclidean graph space. A common strategy in graph change-point detection is to preprocess this trajectory into a low-dimensional Euclidean representation, enabling the application of standard tools from time series analysis and signal processing \cite{lacasa_scalar_2024, huang_laplacian_2020}. Our method follows this philosophy: it first constructs a high-dimensional diffusion-based representation of the temporal network, and then maps it to a low-dimensional signal through local conditional entropy.

The main contributions of this article are fourfold. First, we extend the conditional entropy of heat diffusion from static graphs to continuous-time temporal networks, establish its basic theoretical properties, provide an upper bound on its evolution, and show how it can be used to measure the importance of asymmetric temporal paths. Second, we introduce a local conditional entropy, better suited to capturing structural changes on finite time scales, and study its behavior. Third, we use this quantity to build an entropy-informed graph change-point detection pipeline that applies naturally to both link streams and snapshot networks and achieves competitive, and in some settings superior, performance against established baselines on synthetic benchmarks. To the best of our knowledge, this is the first work to address graph change-point detection directly on link streams with non-zero link durations, rather than first reducing the data to a purely discrete-time snapshot representation. Fourth, we illustrate the practical value of the detected change points on a real-world temporal network by combining them with community detection.
Identifying change points has many downstream applications. For instance, as we show in \cref{sec:Applications}, identifying structurally stable time intervals can improve the interpretability and effectiveness of dynamic community detection.

The paper is organized as follows. In \cref{sec:Theory}, we introduce the theory and notation for temporal networks and heat diffusion over them, and then develop the global and local conditional entropy formalisms. In \cref{sec:NCPD}, we describe the graph-signal processing technique based on local conditional entropy, and evaluate it on synthetic change-point detection benchmarks. In \cref{sec:Applications}, we use a real-world dataset to demonstrate its potential in combination with community detection methods. Finally, we discuss our results, limitations, and future directions in
\cref{sec:DiscussionConclusions}.

\section{\label{sec:Theory} Heat Diffusion, Dynamic Networks, and Conditional Entropy}

We now briefly introduce the notation of diffusion on (dynamic) graphs as a stochastic process. For static graphs, it is a homogeneous Markov Chain; for dynamic graphs, an inhomogeneous one.

\subsection{Heat Diffusion on Networks}
Let $ G = (V, E) $ be an undirected  graph, where
$ V = \{v_1, v_2, \dots, v_N\} $ is the set of $ N  \in \mathbb{N}$ nodes,  and $ E = \{e_1, e_2, \dots, e_M \} \subset V \times V $ is the set of $M \in \mathbb{N}$ edges connecting pairs of nodes.

We model heat diffusion on a graph as a time-homogeneous continuous-time Markov chain $\{ Z(t) \}_{t \in \mathbb{R}^+}$ \cite{masuda_random_2017}, where the state at time $ t $ is $ Z(t) \in  \left\{v_1, v_2, \dots, v_N \right\} $.  We use $p(t)$ to denote the law of $Z(t)$, and $ p_i(t) $ is the probability of $Z(t)$ being equal to $v_i$ at time $t$.
We often manipulate $p(t)$ in vectorized form. We use the bold notation $\p(t)$, and similarly for other vectors (lowercase) and matrices (uppercase).
Heat diffusion is described by the equation
\begin{equation} \label{eq:static_heat_kernel}
    \p^{\intercal}(t) = \p^{\intercal}(0) e^{- \lambda \Lapl t},
\end{equation}
parametrized by the rate of diffusion $\lambda \in \mathbb{R}^+$, and where $\Lapl$ is the combinatorial Laplacian matrix, defined as follows:
\begin{equation} \label{eq:Laplacian}
    \Lapl_{i,j} = \begin{cases} 
      d_i & \text{if } i=j, \\
      - 1 & \text{if } (v_i, v_j) \in E, i \neq j, \\
      0 & \text{otherwise},
   \end{cases}
\end{equation}
where $d_i$ indicates the degree of node $i$. 

Irreducible continuous-time Markov chains are ergodic \cite{lanchier_stochastic_2017}. For heat diffusion this implies that on connected graphs, the probability vector converges to the stationary distribution $ \matr{\pi}^{\intercal} = \frac{1}{N} \matr{1}^{\intercal}_N = [\frac{1}{N}, \frac{1}{N}, \dots, \frac{1}{N}]$, where $\matr{1}_N$ is the $N$-th dimensional all-ones vector.
This stationary distribution $\pi$ reflects the equilibrium state of the diffusion process, where each node’s value remains constant over time, and satisfies the stationarity condition
$\matr{\pi}^{\intercal} = \matr{\pi}^{\intercal} e^{- \lambda \Lapl t}$ for all $t \geqslant 0$.

Several mathematical formalisms can be considered to describe temporal networks. Discrete-time structures, such as snapshot networks \cite{holme_temporal_2012}, are the most commonly studied, particularly for change-point detection. This article focuses on continuous-time dynamic networks known as link streams \cite{latapy_stream_2018}. These are more general than snapshot networks, as the temporal information for each connection is more precise, enabling us to capture finer-grained temporal details and describe richer temporal evolutions. Before extending the definition of diffusion to link streams, we briefly introduce the notation directly related to them, which will be helpful for the remainder of the paper. 

A \emph{link stream} over the node set $V$, within the \emph{time domain} $T$ (an interval of $\mathbb{R}^+$ starting from 0), is a triplet
\begin{equation}
    L=(T, V, E), \qquad 
E \subseteq \{([\alpha,\omega),\{u,v\}) : \alpha<\omega,\ u,v\in V,\ u\neq v\},
\end{equation}
so that $([\alpha,\omega),\{u,v\})\in E$ means the edge/link $\{u,v\}$ is present for all $t\in[\alpha,\omega)$.

For each $t\in T$, the corresponding \emph{instantaneous graph} is
\begin{equation}
    G_t = (V, E_t), \quad
    E_t \coloneqq \left\{\{u,v\} : \exists ([\alpha,\omega),\{u,v\})\in E \text{ such that } t\in[\alpha,\omega)\right\}.
\end{equation}
Given a window $I=[a,b]\subseteq T$, the \emph{unweighted footprint} is the static graph
\begin{equation} \label{eq:footprint_u}
    G_I \coloneqq \left(V,\left\{\{u,v\}: \exists t\in I \text{ such that } \{u,v\}\in E_t \right\}\right).
\end{equation}
Similarly, the \emph{weighted footprint} $\tilde G_I$ assigns to each edge the contact duration $w_I(u,v)\coloneqq \operatorname{Leb}\bigl(\{t\in I : \{u,v\}\in E_t\}\bigr)$.

Heat diffusion can be naturally generalized to link streams. Let
$0=t_0<t_1<\cdots<t_m=\max(T)$
be the temporal grid formed by the ordered set of all link start and end times. By construction, the instantaneous graph remains constant on each interval $[t_k,t_{k+1})$. Diffusion therefore behaves as on a static graph within each such interval. More precisely, if $G_t$ denotes the instantaneous graph at time $t$, and $\Lapl(t)$ its combinatorial Laplacian, then for any interval $[a,b]\subseteq [t_k,t_{k+1})$ we define the corresponding transition matrix by
\begin{equation} \label{eq:inter_T}
    \hat{\T}^\lambda(a,b) \coloneqq e^{-\lambda \Lapl(t_k)(b-a)},
\end{equation}
where $\lambda \in \mathbb{R}^+$ is the rate of diffusion.
In particular, on the full constant interval $[t_k,t_{k+1})$, this gives
$ \hat{\T}^\lambda(t_k,t_{k+1}) = e^{-\lambda \Lapl(t_k)\tau_k}$ where $\tau_k \coloneqq t_{k+1}-t_k$ is an inter-event time.
For two arbitrary times $\bar t_1<\bar t_2$, let $m,n$ be such that
$\bar t_1 \in [t_m,t_{m+1})$, and $\bar t_2 \in [t_n,t_{n+1})$.

Then the transition matrix from time $\bar t_1$ to time $\bar t_2$ is obtained by composition \cite{bovet_flow_2022}:
\begin{equation} \label{eq:inhomogeneous_kernel}
    \T^\lambda(\bar t_1,\bar t_2)
    =
    \hat{\T}^\lambda(\bar t_1,t_{m+1})
    \left(
    \prod_{k=m+1}^{n-1} \hat{\T}^\lambda(t_k,t_{k+1})
    \right)
    \hat{\T}^\lambda(t_n,\bar t_2),
\end{equation}
with the convention that the middle product is omitted when $n\le m+1$. Since we use row-vector convention for probability distributions, the kernels compose from left to right in chronological order: $\T^\lambda(t_1,t_3)=\T^\lambda(t_1,t_2)\T^\lambda(t_2,t_3),
\qquad t_1\le t_2\le t_3$.

\subsection{Conditional Entropy of Heat Diffusion in Temproal Networks} \label{subsec:entropy}
Extending heat-kernel-based entropy measures from static to temporal networks is nontrivial. For example, the spectral entropy proposed in Refs. \cite{de_domenico_spectral_2016,ghavasieh_statistical_2020,villegas_laplacian_2022,villegas_laplacian_2023} is based on the Von Neumann entropy of a density matrix taken as proportional to the heat kernel of the graph, which needs to be positive semidefinite and symmetric. The Von Neumann entropy is then given as the Shannon entropy of its eigenvalues. However, as shown in \cref{appendix:vNS}, in a temporal network, the presence of asymmetric temporal paths induces asymmetries in the heat kernel, which can lead to complex eigenvalues, invalidating the direct extension of the Von Neumann entropy to temporal networks.
We studied the evolution of conditional entropy for diffusion processes on static graphs in Ref. \cite{koovely_evolution_2025}.
We now show that this framework naturally extends to temporal networks. Conditional entropy measures the amount of information needed to describe a random variable once another one is known \cite{cover_elements_2006}.

Let $\{Z(t)\}_{t\in\mathbb{R}^+}$ denote the heat diffusion process on a link stream, and let $p(0)$ be the distribution of $Z(0)$. For each node $i$, we define the row entropy
$$
H_i^\lambda(0,t)
\coloneqq
-\sum_j T^\lambda_{i,j}(0,t)\log\bigl(T^\lambda_{i,j}(0,t)\bigr),
$$
that is, the entropy of the distribution obtained by starting from node $i$ at time $0$ and diffusing until time $t$. The corresponding conditional entropy is then
\begin{equation} \label{eq:conditional_entropy}
    H^\lambda(t, p(0))
    \coloneqq
    H(Z(t)\mid Z(0))
    =
    \sum_i p_i(0)\, H_i^\lambda(0,t)
    =
    -\sum_i p_i(0)\sum_j T^\lambda_{i,j}(0,t)\log\bigl(T^\lambda_{i,j}(0,t)\bigr).
\end{equation}
At time $t=0$, this quantity is equal to $0$ for every choice of initial distribution $p(0)$.

In a static undirected connected network, the Markov chain is time-homogeneous and ergodic. In this case, a related measure is the entropy rate of the Markov chain, defined as its conditional entropy in the stationary state \cite{cover_elements_2006}, which was proposed as an entropic measure for complex networks in Ref. \cite{gomez-gardenes_entropy_2008}. In temporal networks, the Markov chain is generally non-stationary, rendering the generalization of the entropy rate to temporal networks invalid. However, as we show below, the conditional entropy is a well-defined and useful measure.

We showed that, in the static case,  the conditional entropy is non-decreasing in time even for finite-time, before reaching stationarity \cite{koovely_evolution_2025}. This provides an information-theoretic version of the second law of thermodynamics. The same remains true for temporal networks. Namely, for every initial distribution $p(0)$, every diffusion rate $\lambda$, and every $0\le t_1\le t_2$,
\begin{equation} \label{eq:monotonicity_t_global}
    H^\lambda(t_2, p(0))\ge H^\lambda(t_1 , p(0)).
\end{equation}
The proof follows the same argument as in \cite[Thm.3.4]{koovely_evolution_2025}, using the composition property of the inhomogeneous kernels together with the data processing inequality. More precisely, for every node $i$ and every $t_1\le t_2$,
$\T^\lambda_{i,:}(0,t_2)=\T^\lambda_{i,:}(0,t_1)\T^\lambda(t_1,t_2)$,
and therefore
\begin{equation}
    \dkl\!\left(\T^\lambda_{i,:}(0,t_2)\,\middle\|\,\boldsymbol{\pi}\right)
\le
\dkl\!\left(\T^\lambda_{i,:}(0,t_1)\,\middle\|\,\boldsymbol{\pi}\right),
\label{eq:dataproc}
\end{equation}
where $\dkl$ is the Kullback-Leibler divergence between two distributions and $\boldsymbol{\pi}$ denotes the uniform distribution on the node set. The data processing inequality (\cref{eq:dataproc}) states that, over time, the rows of the transition matrix diverge less and less from $\pi$. The conditional entropy averages the behavior of each row, resulting in aggregate monotonicity (\cref{eq:monotonicity_t_global}). 

\begin{figure}
    \includegraphics[width=0.90\linewidth]{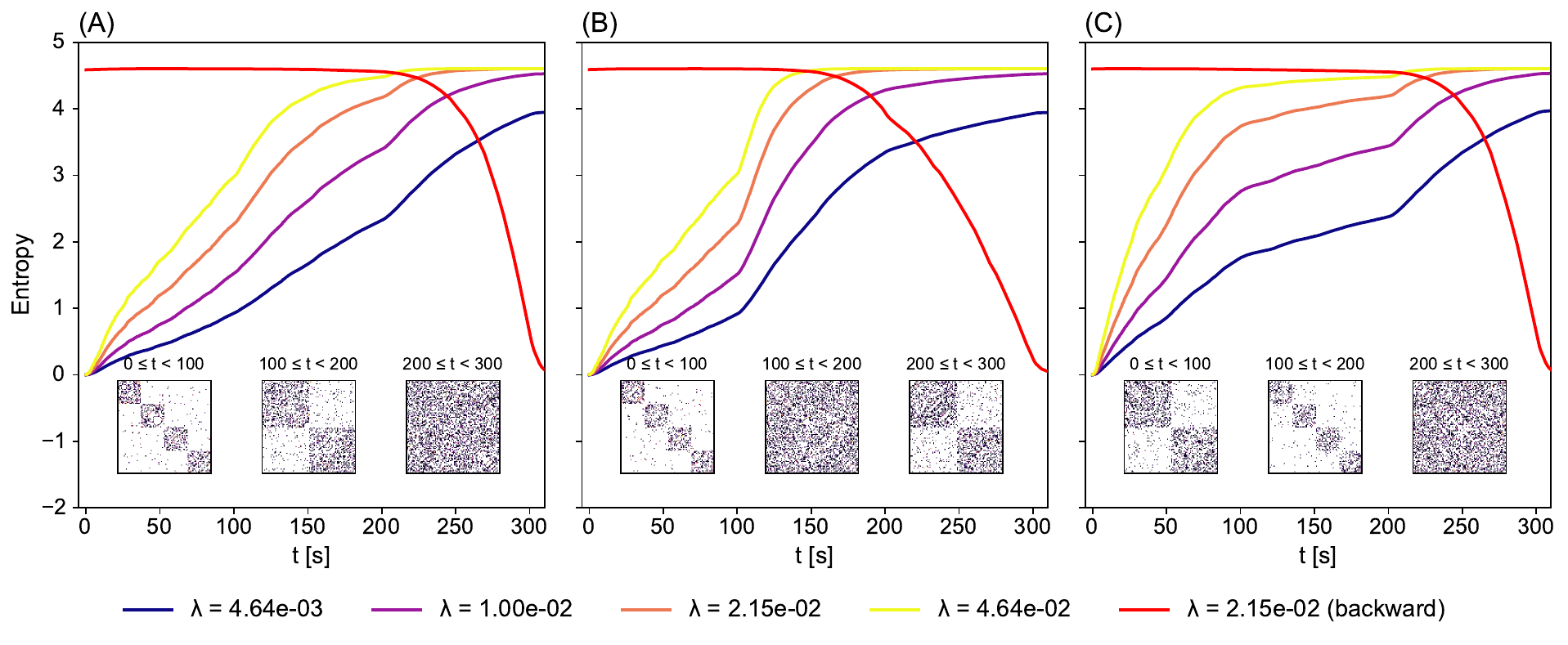}
  \caption{\label{fig:global_entropy} Forward conditional entropy curves for three synthetic temporal networks with 100 nodes experiencing two merge/split events occurring at times 100 and 200. During merges, nodes' activity increases, and they also connect to a broader set of neighbors; during splits, the opposite occurs. Insets show the evolution of the community structure with the adjacency matrices of weighted footprints. (A) In this example, the network incurs two merges: initially, there are 4 clusters of 25 nodes each. During the first event, clusters merge in pairs, resulting in a two-block structure. Later, these two blocks merge as well. 
  The forward entropy increases in time and jumps at the two change points as the structure changes. Depending on the choice of the diffusion rate, these changes are more or less accentuated. The backward curve (in red) increases backward in time and does not observe any jump as it discovers the densest and broadest topology immediately.
  (B) This network incurs first a merge and then a split, going from 4 communities first, to one between times 100 and 200, and finally two. In this case, the forward curves experience only one jump at time 100, whereas the backward entropy jumps at time 200. (C) The forward entropy identifies the merge event at time 200; the backward entropy behaves as in (A). The change at time 100 is missed by both forward and backward entropy because the intermediate scale is smaller than the one at the two extremities.}
\end{figure}

Figure \ref{fig:global_entropy} illustrates the monotonicity property of the conditional entropy on three synthetic temporal networks with $100$ nodes, each undergoing two merge/split events and involving either $1$, $2$, or $4$ communities. During merges, nodes' activity increases, and they also connect to a broader set of neighbors; during splits, the opposite occurs.
In \cref{fig:global_entropy}, the entropy is non-decreasing, and sudden increases in its rate may signal the presence of merge events, because first the number of non-zero entries in the heat kernel decreases, and more generally, the increase in connectivity brings the rows closer to the uniform distribution $\T^\lambda_{i,:}(0, t)$. Because time-respecting paths are generally asymmetric in temporal networks, it is also natural to study the conditional entropy associated with the time-reversed evolution \cite{bovet_flow_2022}.
Consider the heat kernel obtained by applying the same construction to the temporal network reversed in time. The corresponding backward entropy satisfies the same monotonicity property on the reversed network; when re-expressed in the original time direction, it therefore becomes non-increasing and may reveal split events through a sudden decrease in rate. Depending on the choice of the diffusion rate, these jumps or dips are more or less pronounced: for low values of $\lambda$, diffusion may be too slow to detect the structural change, whereas for large values it may saturate too quickly, making subsequent changes difficult to detect. Given that both forward and backward diffusion processes are non-reversible, they do not react sharply to the appearance of finer-scale structures where diffusion slows down, as pointed out in panel (C) of \cref{fig:global_entropy}.

\begin{figure}[htbp]
    \centering
    \includegraphics[scale=0.75]{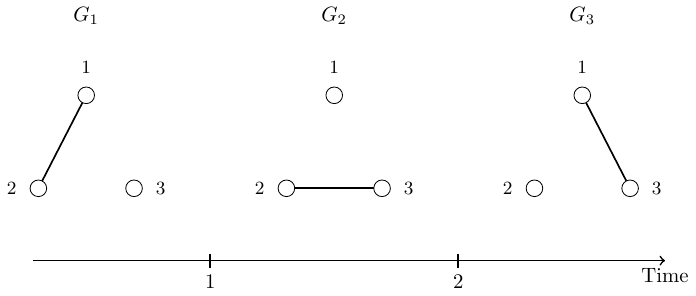}
      \caption{\label{fig:Counterexample} Temporal cycle with three nodes. Nodes 1 and 2 are connected from time 0 to 1, nodes 1 and 2 from time 1 to 2, and nodes 1 and 3 from time 2 to 3.}
\end{figure}

The figure seems to suggest that monotonicity with respect to the diffusion rate holds as well; however, this is not true in general. This is slightly counterintuitive: between changes, increasing $\lambda$ increases mixing over the same interval, the equivalent of increasing the diffusion time in static networks. However, this local behavior does not hold globally, and a counterexample is illustrated in \cref{fig:Counterexample}, a temporal cycle with three nodes. In that case, if we take $p(0)$ to be the uniform distribution, one can verify that  $H^{1}(3, p(0)) > H^{10}(3, p(0))$.

The behavior of the conditional entropy on temporal networks is better understood by first considering it on static networks. In static connected graphs, heat diffusion converges to the uniform stationary distribution, and the conditional entropy converges to its maximal value $\log(N)$. For disconnected graphs, the asymptotic value depends only on the connected-component sizes \cite[Prop.3.10]{koovely_evolution_2025}: if $V=\bigcup_{k=1}^\ell V_k$ is the decomposition into connected components and the initial distribution is uniform, then
\begin{equation} \label{eq:limit_curve}
    \lim_{t\to\infty} H^\lambda(t, p(0))
    =
    \sum_{k=1}^\ell \frac{|V_k|}{N}\log(|V_k|),
\end{equation}
where $p(0)$ is the uniform distribution on $V$.
This formula relies on the symmetry of the transition matrices and the existence of an asymptotic distribution. Temporal graphs, by contrast, are non-stationary and their heat kernels are generally not symmetric, so no exact analog holds in full generality.

We therefore obtain a natural upper bound by applying \cref{eq:limit_curve} to the connected components of the static footprint $G_{[0,t]}$.
We have shown that the entropy is not monotone in $\lambda$ in full generality, but in practice, simulations suggest that over complex networks, approximately, it is. We therefore, conjecture that the limit curve $\lim_{\lambda\to\infty} H^\lambda(t, p(0))$ is well defined. This limit may be much smaller than the above-mentioned upper bound due to the diffusion kernel's temporal asymmetry. This difference between the limiting values of the conditional entropy on a temporal network and on its static aggregation provides a measure of the extent to which time-respecting paths constrain diffusion, as discussed in more detail now.

\subsection{Bounds and Temporal Asymmetry}

\begin{figure}[htbp]   \includegraphics[width=0.95\linewidth]{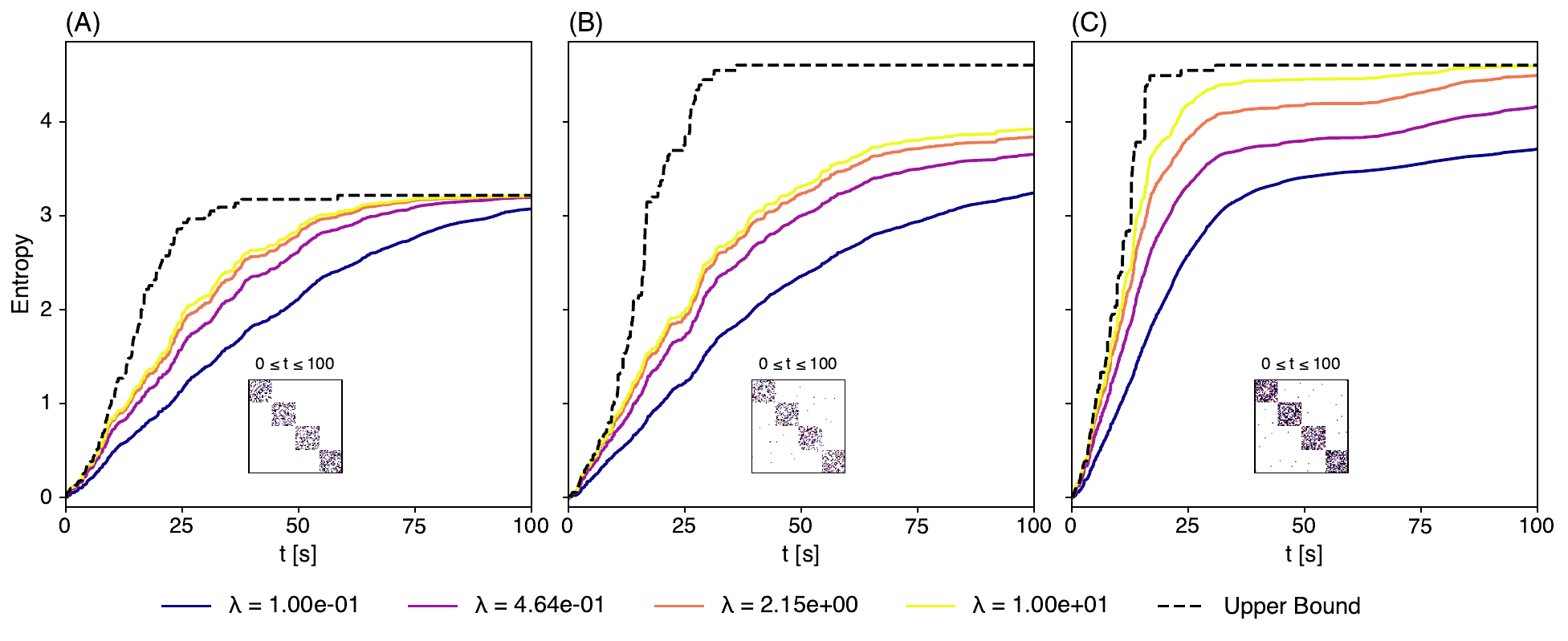}
    \caption{\label{fig:fig_global_entropy_edlde_upper_bound} Comparison between upper bound estimate and forward entropy curves in three synthetic temporal networks with 100 nodes. All networks have four communities, but different connectivity patterns. (A) Links appear with a frequency of 5 Hz, and there are no cross-community edges. 
  Despite the low density, the asymptotic curve reaches the saturation level towards the end.
  (B) Networks with exact properties as in (A), but with some spurious cross-community links. Their presence raises the upper bound estimate, which the entropy curve does not reach, even as we approach the end time of 100s, and when the asymptote in diffusion rate is reached. This is because the network's density is low, so there are not enough links that reappear multiple times or create circles, compensating for the presence of asymmetric temporal paths. (C) Networks with exact properties as in (B), but links appear with a frequency of 10 Hz. Given the higher density, there are more symmetric time-respecting paths to keep the asymptotic curve close to the upper bound.}
\end{figure}

In the previous part, we derived an upper bound for conditional entropy curves by applying \cref{eq:limit_curve} to the connected components of the static footprint $G_{[0,t]}$. The actual limit may be much smaller due to the diffusion kernel's temporal asymmetry, as shown in \cref{fig:fig_global_entropy_edlde_upper_bound}. All panels show conditional entropy curves on temporal networks with static structures lasting 100s. The network considered in panel (B) has the same properties as the one from panel (A), with the addition of some spurious cross-community links. Their presence raises the upper-bound estimate, so the entropy curves do not reach it even at the end time 100s. If we raise the spawning rate of links, as shown in panel (C), we create additional paths that compensate for earlier asymmetries of the heat kernel. As a consequence, the entropy curves are closer to the upper bound compared to those from panel (B).

Thus, the difference between the limiting values of the conditional entropy, in terms of the rate of diffusion, on a temporal network and on its static aggregation provides a measure of the extent to which time-respecting paths constrain diffusion. We clarify this point with a deterministic toy example where the number of asymmetric time-respecting paths can be controlled. We consider a family of progressively synchronized temporal paths that
interpolates between a fully ordered temporal path and a fully synchronized one, while keeping the underlying unweighted footprint fixed.

For each integer $N \geq 2$ and each \emph{overlap depth}
$k \in \{0,\dots,N-2\}$, we define the link stream
\begin{equation*}
L_{N,k} = (T_N,V_N,E_{N,k}),
\qquad
T_N = [0,N-1),
\qquad
V_N = \{1,\dots,N\}.
\end{equation*}
Its unweighted footprint $\tilde{G}_{T}$ is the path on $N$ nodes, i.e., the edges are
$\{i,i+1\}$ for $i=1,\dots,N-1$.
The parameter $k$ controls how many of the first links share the same temporal support. More precisely, for each $i \in \{1,\dots,N-1\}$, we set
\begin{equation}
    E_{N,k}
    =
    \left\{
    \bigl(I_i^{(k)},\{i,i+1\}\bigr)
    :\,
    i=1,\dots,N-1
    \right\}, \text{where }
    I_i^{(k)} =
    \begin{cases}
    [0,k+1), & \text{if } i \leq k+1,\\
    [i-1,i), & \text{if } i \geq k+2.
    \end{cases}
\end{equation}
Thus, when $k=0$, all links are temporally disjoint:
\begin{equation*}
E_{N,0}
=
\left\{
([0,1),\{1,2\}),
([1,2),\{2,3\}),
\dots,
([N-2,N-1),\{N-1,N\})
\right\}.
\end{equation*}
When $k=1$, the first two links overlap completely on $[0,2)$, while the remaining ones keep their original disjoint supports. At the opposite extreme, when $k=N-2$, all $N-1$ links are present on the common interval
$[0,N-1)$, so that
\begin{equation*}
E_{N,N-2}
=
\left\{
([0,N-1),\{i,i+1\})
:\,
i=1,\dots,N-1
\right\}.
\end{equation*}

Hence, increasing $k$ reduces the temporal asymmetry of the path by merging the supports of an increasing prefix of links. In particular, for large diffusion rates $\lambda$, the final global entropy is expected to increase
with $k$ and to approach the upper bound $\log N$ as $k \to N-2$.

For instance, for $N=4$, the three cases are
\begin{equation*}
L_{4,0}:\quad
([0,1),\{1,2\}),\ 
([1,2),\{2,3\}),\ 
([2,3),\{3,4\}),
\end{equation*}
\begin{equation*}
L_{4,1}:\quad
([0,2),\{1,2\}),\ 
([0,2),\{2,3\}),\ 
([2,3),\{3,4\}),
\end{equation*}
and
\begin{equation*}
L_{4,2}:\quad
([0,3),\{1,2\}),\ 
([0,3),\{2,3\}),\ 
([0,3),\{3,4\}).
\end{equation*}

\begin{figure}[htbp]   \includegraphics[width=0.70\linewidth]{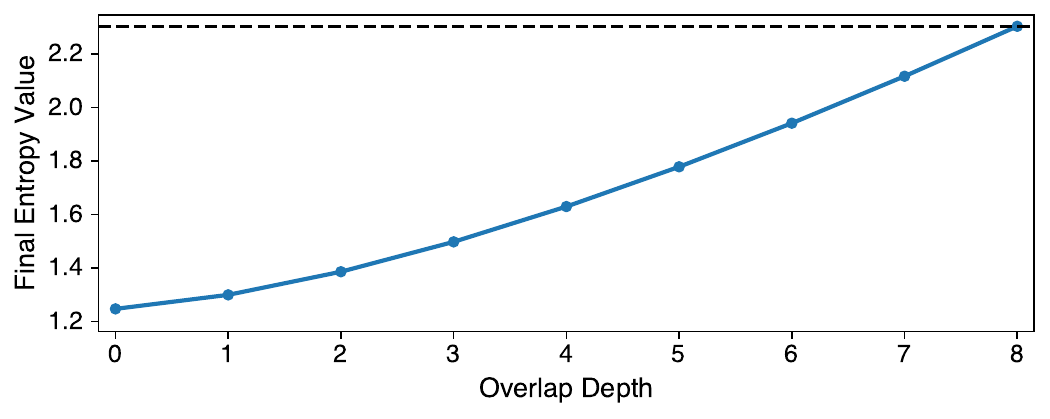}
\centering
    \caption{\label{fig:toy_path_final_global_entropy} Comparison between the final global entropy value for $L_{10, k}$ for $k \in \{0, \dots, 8\}$ and the upper bound $\log(10)$, depicted with a dashed horizontal black line. This toy example shows that as we increase the symmetry of the heat kernel, the conditional entropy approaches the upper bound given by the conditional entropy on the aggregated static network.}
\end{figure}

We capture the effect of asymmetry on the tightness of the upper bound in \cref{fig:toy_path_final_global_entropy}, where, for $k \in \{0, \dots, 8\}$, we compare the final global entropy values for $L_{10, k}$ to the upper bound. As we merge more supports, the heat kernel becomes more symmetrical by construction, resulting in a final entropy value closer to the bound.

The case $k=0$ can be studied explicitly and provides a useful reference
value. 
The fully ordered temporal path, $L_{N,0}$, is the temporal network with a connected footprint that has the least number of edges and also has the least number of symmetric temporal paths, since edges have no temporal overlap. In this sense, it is a candidate for a lower bound on the evolution of the conditional entropy in connected networks of size $N$.
In this case, the links of the temporal path appear one after another, with no temporal overlap:
\begin{equation*}
E_{N,0}
=
\left\{
([0,1),\{1,2\}),
([1,2),\{2,3\}),
\dots,
([N-2,N-1),\{N-1,N\})
\right\}.
\end{equation*}
Let $T^\lambda_{N,0}$ denote the transition matrix of heat diffusion on $L_{N,0}$ from time $0$
to time $N-1$. We consider its asymptotic value in the diffusion rate, $T^\infty_{N,0} \coloneqq \lim_{\lambda \to \infty} T^\lambda_{N,0}$.
Since at each time point the network consists only of isolated vertices and one active edge, the limit $\lambda \to \infty$ has a simple interpretation: whenever a link $\{i,i+1\}$ is active, the mass on $i$ and $i+1$ is instantaneously redistributed uniformly over the pair $\{i,i+1\}$, while all other vertices remain unchanged.

In \cref{appendix:path_graph} we show that if $Z(0)$ is uniformly distributed over $V_N$, the limiting final conditional entropy of $L_{N,0}$ is
\begin{equation}
    H(T^\infty_{N,0}\mid Z(0)) = \left(2-\frac{2}{N}\right)\log(2).
    \label{eq:L_N_0}
\end{equation}

This shows explicitly that, although the static footprint is connected, the
temporal ordering of the links can keep the asymptotic entropy much smaller
than $\log(N)$. In particular, for large $N$, the fully ordered temporal path has limiting entropy close to $2\log(2)$, rather than growing like $\log(N)$. 
This result allows us to formulate a candidate lower bound for the evolution of the conditional entropy on temporal networks.
Namely, if $V=\bigcup_{k=1}^\ell V_k$ is the decomposition into connected components of the static footprint $G_{[0,t]}$, and the initial distribution is uniform, we conjecture that the asymptotic curve of the conditional entropy has a lower bound given by
\begin{equation} \label{eq:lower_bound}
    \sum_{k=1}^\ell \frac{|V_k|}{N}\left(2-\frac{2}{|V_k|}\right)\log(2).
\end{equation}

We illustrate the upper and lower bounds of the conditional entropy in Fig. \ref{fig:primaryschool_fastest_bounds} in Appendix \ref{appendix:path_graph}.

\subsection{Local Conditional Entropy}
The conditional entropy measures changes in the row-probability distributions of the transition matrix of a process on a temporal network and is therefore a potential measure of structural change. However, as illustrated in \cref{fig:global_entropy}, its temporal monotonicity implies that it is not strongly affected by structural changes that restrict diffusion, such as community splits, and ultimately reaches a limit. 
To quantify changes in diffusion pathways at a local temporal scale, independent of the full past history of the process, we therefore introduce a local version of the conditional entropy for a time window $\Delta$. For $\frac{\Delta}{2}\le t \le \max(T)-\frac{\Delta}{2}$, we define
\begin{equation} \label{eq:loc_conditional_entropy}
    H(t,\lambda,\Delta)
    \coloneqq
    -\sum_i \frac{1}{N}\sum_j
    T^\lambda_{i,j}\!\left(t-\frac{\Delta}{2},t+\frac{\Delta}{2}\right)
    \log\!\left(
    T^\lambda_{i,j}\!\left(t-\frac{\Delta}{2},t+\frac{\Delta}{2}\right)
    \right).
\end{equation}
Equivalently, $H(t,\lambda,\Delta)$ is the conditional entropy associated with the kernel from $t-\Delta/2$ to $t+\Delta/2$, with uniform initial distribution at time $t-\Delta/2$.

The monotonicity principle derived for the global conditional entropy implies that the local entropy is still non-decreasing with respect to the window length. More precisely,
\begin{equation} \label{eq:monotonicity_Delta_local}
    \forall t,\ \forall \lambda,\ \forall \Delta_1\le \Delta_2,
    \qquad
    H(t,\lambda,\Delta_2)\ge H(t,\lambda,\Delta_1).
\end{equation}
Indeed, increasing $\Delta$ enlarges the diffusion interval and therefore can only increase the amount of diffusion captured by the kernel.

\begin{figure}[htbp]
    \includegraphics[width=0.95\linewidth]{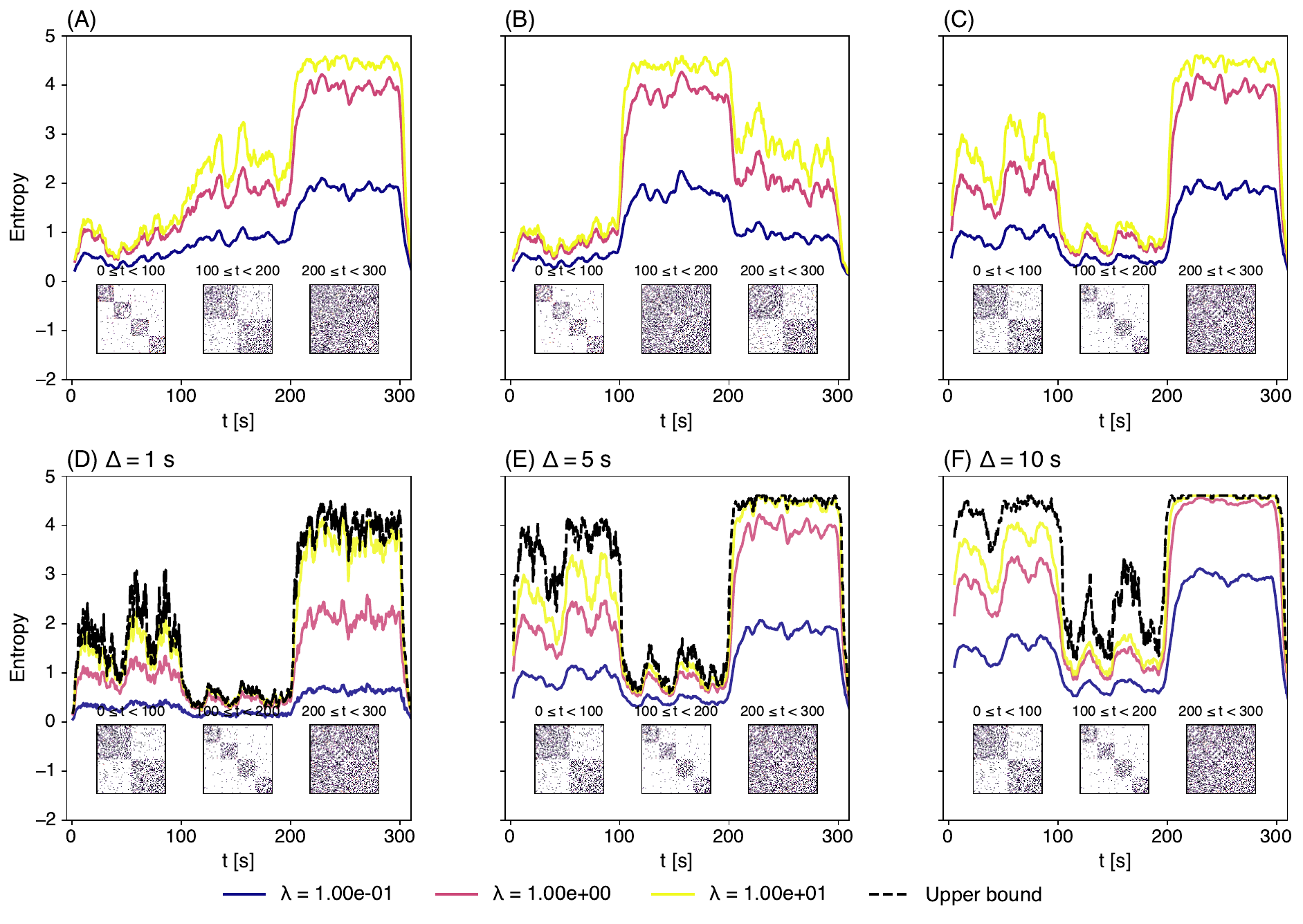}
  \caption{\label{fig:local_entropy_combined} The three panels on top, show local entropy curves $H(t, \lambda, 5)$ for various diffusion rates $\lambda$ over the same temporal networks discussed in \cref{fig:global_entropy}. Each panel contains forward curves for different diffusion rates. Unlike the global entropy, the structural changes are clearly visible in all cases. The bottom panels compare entropy curves across different time windows for the third temporal network analyzed in \cref{fig:global_entropy}, shown here in panel (C). We include upper limit curves according to \cref{eq:limit_curve} as dashed black lines. The bound is now tighter compared to the global entropy, especially for smaller time windows, or when there are sufficient links to bring the signal close to saturation.}
\end{figure}

On the other hand, the local conditional entropy is no longer monotone in time; instead, it behaves as a local measure of diffusion-induced mixing. As in the global case, the limit $\lim_{\lambda\to\infty} H(t,\lambda,\Delta)$ appears to be well defined. A natural upper bound is obtained by applying \cref{eq:limit_curve} to the connected components of the footprint $G_{[t-\frac{\Delta}{2},\,t+\frac{\Delta}{2}]}$.
In \cref{fig:local_entropy_combined}, we show the behavior of the local entropy on the same synthetic temporal networks as in \cref{fig:global_entropy}. Given that the structural information is more local, the discrepancy between the forward and backward processes is very small; henceforth, we discuss only the forward process.
The top panels display the entropy signal $H(t, \lambda, 5)$ for various values of the diffusion rate $\lambda$, in all three cases of pairs of merge/split events discussed in \cref{fig:global_entropy}. The signals appear to be approximately piecewise linear, with some variance due to stochastic fluctuations. The structural changes are more transparent compared to the corresponding global conditional entropy ones seen previously, even in the split-merge example.
In the bottom panels, we focus on this last example, and compare the signals for three different windows $\Delta$ of $H(t, \lambda, \Delta)$. Increasing the window size smooths the signal but can saturate it, as shown in panel (F), where the initial phase has an entropy value close to that of the last phase. We also included the upper limit curves based on \cref{eq:limit_curve} as a reference. The figure shows that the bound is now tighter than in the global entropy case, particularly for small time windows. 

\section{Graph Change-Point Detection}\label{sec:NCPD}
Networks are characterized by their structure, and temporal networks by the evolution of that structure. Community evolution typically follows archetypal changes \cite{cazabet_challenges_2023}, often called \emph{events} (such as births, deaths, merges, and splits of communities), separated by change points. Informally, the graph change-point detection task consists of identifying, as accurately as possible, the set of change points $\mathcal{T} = \{ \tau_1, \dots, \tau_K \}$ of a temporal graph.
One distinguishes between the offline and online versions of the problem. In the offline setting, change points are detected a posteriori, after the full graph time series has been observed. In the online setting, one aims to detect a change as soon as new graph observations become available, possibly with some delay.

In this article, we address the offline problem. Formally, let $\{ G_t \}_{t \in \mathbb{R}^+}$ be a family of instantaneous graphs, together with an unknown number $K \in \mathbb{N}$ of change points $\tau_1 < \dots < \tau_K$, such that:
\begin{equation} \label{eq:changepoint}
    \forall k \in \{1, \dots, K \},  \forall t  \in [\tau_{k-1},  \tau_k): G_t \sim \mathcal{G}_{k-1}, 
\end{equation} where $\tau_0 \coloneqq 0$, and $\mathcal{G}_0, \dots, \mathcal{G}_{K-1}$ are graph-generating models associated with the successive regimes.

Following Refs. \cite{huang_laplacian_2020, wang_fast_2017}, we distinguish change points from anomalies: a change point marks a persistent modification of the underlying generative process, whereas an anomaly corresponds to a temporary deviation from the expected behavior. In this work, we focus on change points.

We now describe how local conditional entropy can be used to process temporal networks and perform offline change-point detection.
A temporal network may be viewed as a trajectory in an abstract graph space. A common strategy in graph change-point detection is to preprocess this trajectory into a low-dimensional representation, so that classical tools from time series analysis and signal processing can then be applied; see \cite{lacasa_scalar_2024, huang_laplacian_2020} and \cref{subsec:segmentation}. In our setting, the heat diffusion process provides a high-dimensional representation of both static and temporal networks, encoding structural information through the eigenspaces and spectrum of the corresponding diffusion operators \cite{huang_laplacian_2020, huang_laplacian_2024, blaskovic_random_2025}. The local conditional entropy then yields a low-dimensional summary of this representation that remains sensitive to the underlying information flows. The parameters $\lambda$ and $\Delta$ may be interpreted as hyperparameters that control the smoothness and the structural signal-to-noise ratio of the resulting signal.

Concretely, let $L=(T,V,E)$ be a link stream, and let
$0=t_0<t_1<\cdots<t_m=\max(T)$ denote the times at which the instantaneous graph changes. For fixed parameters $\lambda$ and $\Delta$, we define the discrete signal
\begin{equation}
    y_i \coloneqq H(t_i,\lambda,\Delta),
\end{equation}
for all indices $i$ such that $\frac{\Delta}{2} \le t_i \le \max(T)-\frac{\Delta}{2}$.
We expect the behavior of the resulting signal $y=\{y_i\}$ to reflect the evolution of the underlying temporal network, so that graph change points in $L$ correspond to numerical change points in $y$. In this way, the graph change-point detection problem reduces to a classical change-point detection or segmentation problem.

\subsection{Segmentation} \label{subsec:segmentation}
We now describe the segmentation task in more detail, following loosely the notation of \cite{truong_selective_2020}.
Let $y=\{y_t\}_{t=1}^{\Gamma}$ be a signal of length $\Gamma$. For integers $0 \le a < b \le \Gamma$, we denote by $y_{a..b} \coloneqq \{y_t\}_{t=a+1}^{b}$
the sub-signal consisting of the observations with indices $a+1,\dots,b$. In particular, the full signal is written $y=y_{0..\Gamma}$.
We assume that the signal is piecewise stationary, meaning that its probabilistic behavior remains stable over intervals and changes abruptly at unknown times $\tau_1<\cdots<\tau_K$.
The goal of segmentation is to estimate the change point set
$\hat{\mathcal T}=\{\hat\tau_1,\dots,\hat\tau_{\hat K}\}$, and in some situations also the unknown number $K$ of changes.

A standard approach is to formulate segmentation as a model selection problem. One introduces a criterion $V(\hat{\mathcal T},y)$ evaluating the quality of a candidate segmentation, and seeks a segmentation minimizing this quantity. In the methods considered here, the criterion is assumed to decompose additively over segments:
\begin{equation}
V(\hat{\mathcal T},y)=\sum_{k=0}^{\hat K} c\!\left(y_{\hat\tau_k..\hat\tau_{k+1}}\right),
\label{eq:additive_cost}
\end{equation}
where $\hat\tau_0 \coloneqq 0$, $\hat\tau_{\hat K+1} \coloneqq \Gamma$, and $c(\cdot)$ is a segment cost function measuring how well a given sub-signal fits a chosen model. Segments containing no internal change should typically produce a small cost, whereas segments containing one or more changes should yield a larger value.

Two standard optimization problems arise. If the number of change points is fixed in advance, one solves
\begin{equation}
\min_{|\hat{\mathcal{T}}|=K} V(\hat{\mathcal{T}}, y).
\tag{P1}
\end{equation}
If the number of changes is unknown, one instead adds a penalty term $\mathrm{pen}(\hat{\mathcal T},y)$ controlling model complexity and avoiding over-segmentation. One then solves
\begin{equation}
\min_{\hat{\mathcal{T}}} \left\{ V(\hat{\mathcal{T}}, y) + \mathrm{pen}(\hat{\mathcal{T}}, y) \right\}.
\tag{P2}
\end{equation}

Once a criterion $V$ has been chosen, one must solve the associated optimization problem, either exactly or approximately. Different search strategies trade off statistical accuracy against computational efficiency.

\subsection{Entropy-Informed Graph Change-Point Detection}
\begin{figure}[htbp]
    \includegraphics[width=0.95\linewidth]{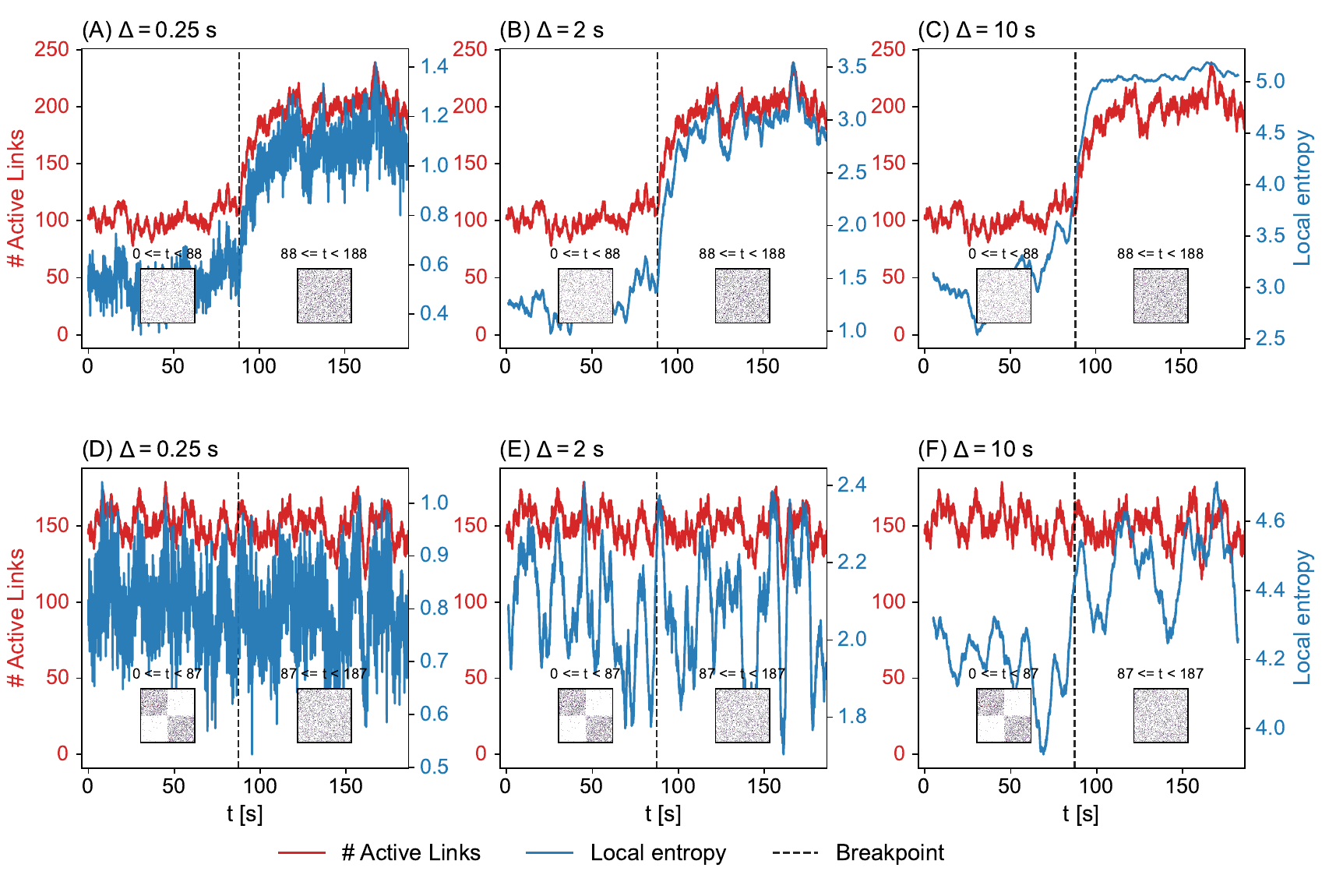}
\caption{\label{fig:ct_examples_combined}  On the top row, we study one example of a network lasting a bit less than 200s with 200 nodes without a structure, where the average number of links changes drastically at 88s from the start. We compute and show the entropy signals $H(t, 1,\Delta)$, for $\Delta \in \{0.25, 2, 10 \}$, which capture the evolution in network activity. Qualitatively, for small context windows, the signal closely follows the count of active links; larger windows are smoothed versions of that curve. We display entropy signals  $H(t, 1,\Delta)$, for $\Delta \in \{0.25, 2, 10 \}$, for a network of 200 nodes with stationary activity, where the community structure changes at 87s. The bottom row's figure highlights that one needs to use a sufficiently wide context window to catch this type of change.}
\end{figure}

Structural changes in real-world temporal networks typically combine changes in network activity (number of active links) and the evolution of communities (increases and decreases in sizes and rewiring mechanisms). An ideal method for network change-point detection should therefore work for both changes that do not involve variation in activity and those that do. We showcase that the local conditional entropy highlights both types of changes in \cref{fig:ct_examples_combined}.
From the top row, one can also see that, in a network without structure, where all that matters is node activity, this activity is captured by the entropy signal. Qualitatively, for small context windows, the signal closely follows the count of active links; larger windows are smoothed versions of that curve.
The second row highlights that one uses a wide context window to catch the second type of change: for small windows, the signal is dominated by fluctuations of the network's activity, but a sufficiently large $\Delta$ allows one to distinguish between the different community structures.

\begin{figure}[htbp]
  \includegraphics[width=0.95\linewidth]{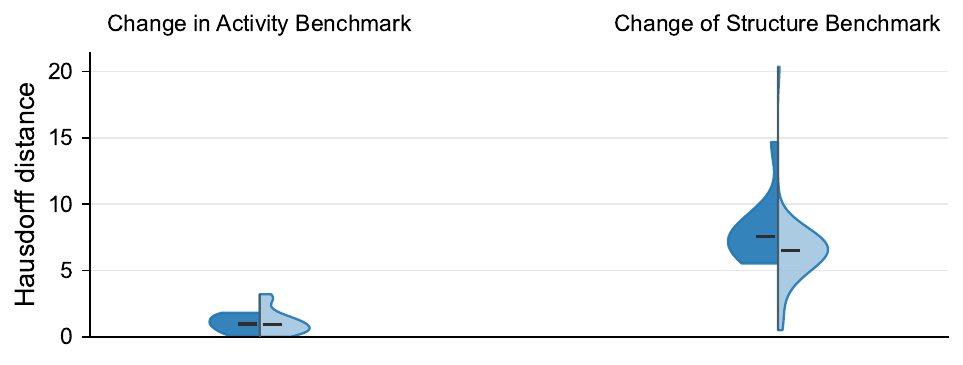}  \caption{\label{fig:fig_ct_experiments_train_test_split_violins} Comparison of performance of the entropy method to detect the two types of changes. We display the distributions of the Hausdorff metric for each sample using split violin plots, with the training sets on the left and the test sets on the right of each violin. The pure change of communities is detected with a larger 20s windows, therefore with less precision.}
\end{figure}

In the case a classic training-test set-up for this kind of tasks, one only needs a minimal number of training points to select, an appropriate pair of parameters $\lambda$ and $\Delta$ to calibrate the method.
We evaluate our method on both task types using two benchmarks: a test set of 50 networks and a training set of 10 networks for tuning the parameters $\lambda$ and $\Delta$. 
Several metrics have been proposed to evaluate the quality of a segmentation. We use the Hausdorff metric \cite{boysen_consistencies_2009, harchaoui_multiple_2010, truong_selective_2020}, which penalizes both over- and under-segmentation, and is appropriate even for the case when the number of changes is unknown,
\begin{equation}
    \operatorname{HAUSDORFF}(\mathcal{T}, \hat{\mathcal{T}})
    \coloneqq
    \max \left\{
    \max_{\hat{\tau} \in \hat{\mathcal{T}}} \min_{\tau \in \mathcal{T}} |\hat{\tau} - \tau|,
    \;
    \max_{\tau \in \mathcal{T}} \min_{\hat{\tau} \in \hat{\mathcal{T}}} |\tau - \hat{\tau}|
    \right\}.
\end{equation}
In the special case where $|\mathcal{T}| = |\hat{\mathcal{T}}| = 1$,  the Hausdorff metric becomes just the distance between the predicted and true change point.
The results are shown in \cref{fig:fig_ct_experiments_train_test_split_violins}. Both types of change are detectable, but pure changes in community structure while the network density remains constant require a larger context window $\Delta$ and are therefore detected with lower precision.

To perform these tests, we relied on smooth stochastic block models  \cite{koovely_generating_2025}, which generalize stochastic block models (SMBs) \cite{abbe_community_2018} to continuous-time networks, allowing the generation of smooth structural changes. However, the most commonly studied framework for temporal networks is the snapshot one. Conceptually, we can think of snapshots as being obtained by discretization of the continuous-time domain: we can therefore obtain them from smooth SMBs by splitting the time domain into slices of length $w$ and then building a graph for each time slice, whose nodes and links represent the interactions that occurred during this time slice. 
Since most methods for network change-point detection are designed for snapshot networks, we now expand our experimental discussion to include snapshot networks.

We briefly remark that dynamic SBMs are an alternative for sampling snapshot networks with block structure and have previously been used to test various change-point detection methods \cite{bhattacharjee_change_2020, sulem_graph_2024}.
However, the dynamic SBMs' snapshots networks have an usually undesired property: all graphs are sampled independently from each other, so there are no correlations between links' presence, even for close snapshots. 

\subsection{Comparative Tests}
 There are several types of approaches to graph chang-point detection \cite{ranshous_anomaly_2015}. Here, we focus on methods that are close in
spirit to our own, namely methods that turn a graph sequence into a
time-indexed signal and then detect changes from that signal. We therefore do not consider approaches based on explicit generative statistical models or supervised neural architectures, since these require, typically, substantially larger training sets.
 
Following an experimental setup close to that of
\cite{sulem_graph_2024}, we compare our method with two baselines that are also
used there: a direct matrix-distance method based on the Frobenius norm \cite{barnett_change_2016}, and a
spectral method based on Laplacian Anomaly Detection (LAD)
\cite{huang_laplacian_2020,huang_laplacian_2024}. 

Although graph change-point detection does not yet have a universally agreed-upon state of the art, these two baselines are representative of common strategies for processing snapshot network sequences.

The Frobenius baseline works directly on the snapshot matrices, without introducing a spectral embedding \cite{barnett_change_2016, sulem_graph_2024}. For each time $t$, it compares the current adjacency matrix $A_t$ with the matrices in a backward window and averages the corresponding normalized Frobenius distances:
\begin{equation}
    F(t) \coloneqq \frac{1}{\ell}\sum_{j=1}^{\ell}
\frac{\lVert A_t - A_{t-j}\rVert_F^2}{\lVert A_t\rVert_F\,\lVert A_{t-j}\rVert_F}.
\end{equation}

Large values of $F(t)$ indicate that the current graph differs substantially from its recent past, so peaks in this signal can be interpreted as candidate change-points.

LAD represents each graph snapshot $H_t$ through the spectrum of its Laplacian $L_t$ \cite{huang_laplacian_2020,huang_laplacian_2024}. In practice, the method extracts a low-dimensional signature $\sigma_t$ from the leading singular values of $L_t$, which summarizes the global structure of the network. The current signature is then compared with the dominant direction of the signatures observed over a recent sliding window, producing an anomaly score. High values indicate that the present snapshot departs from recent normal behavior; short- and long-term windows can also be combined to capture both abrupt changes and slower drifts.

As mentioned above, both comparison methods are naturally suited to snapshot
networks. If we denote snapshots by $\{H_t \}_{t \in \mathbb{N}}$, then
change-points can be defined, analogously to \cref{eq:changepoint}, as the
delimiting times of piecewise stable regimes:
\begin{equation} \label{eq:snapshotsnet}
    \forall k \in \{ 1, \dots, K \}, \forall t \in  \{\tau_{k-1}, \tau_{k-1} + 1,  \dots, \tau_k -1 \}: H_t \sim \mathcal{H}_{k-1}, 
\end{equation} where $(\mathcal{H}_0, \dots, \mathcal{H}_{K-1})$  denotes the successive snapshot-generating models.
In this setting, our entropy approach consists of taking the signal $H(t, \lambda, \Delta)$ computed over the snapshot sequence, with the natural choice $\Delta = \frac{w}{2}$ (where $w$ is the time-length of each snapshot). This choice allows the context window $\Delta$ to capture structural information within each aggregation interval, without mixing information from adjacent snapshots.

All the experiments in this subsection are obtained by aggregating the
underlying link stream networks with time domain close to $[0,200]$ into non-overlapping binary snapshots of width
$w=4$, hence obtaining samples with around 50 snapshots. Ground-truth breakpoints are then projected from continuous time to snapshot time by integer division by $w$, so that both detection and
evaluation are carried out directly in snapshot index space. We consider three
benchmark families: \emph{Benchamrk1}, \emph{CommunityBench}, and \emph{MultiBench}, all containing $10$ training samples and $50$ test samples.

The first two benchmarks correspond to the single-change setting. In both
cases, each sample contains exactly one change-point, and the number of changes is assumed to be known. Hyperparameters are selected on a training set by
minimizing the mean Hausdorff distance. For the entropy
method, we tune $\lambda$ over a logarithmic grid in $[10^{-5},1]$, while
keeping the snapshot-scale window fixed at $\Delta=w/2=2$. For the
Frobenius baseline, we tune the backward window length. For LAD, we tune both
the number of retained spectral components and the sliding-window length used
to build the anomaly score.

The \emph{ActivityBench} benchmark contains networks where the structural
model is kept fixed, while the activity changes once between two regimes of
different interaction intensity. The task is therefore to detect a change in temporal activity. This setting is informative because it separates methods that are mainly sensitive to rate variations from those that remain effective when the signal is driven by broader structural changes.

The \emph{CommunityBench} benchmark isolates a structural change under approximately constant activity. The overall interaction rate remains essentially unchanged, while the latent block organization is modified once during the observation interval. In other words, the detector cannot rely on a simple global increase or decrease in the number of interactions; it must instead react to a change in the underlying block organization. This makes \emph{CommunityBench} a useful stress test for methods that aim to detect mesoscopic reorganization rather than only density fluctuations.

We then turn to the more difficult \emph{MultiBench} benchmark, which extends the \emph{ActivityBench} construction to repeated changes. In each sample, the activity alternates between the same two regimes as above, but now the number of breakpoints is itself random, ranging from one to four. This produces a piecewise-stationary activity profile with several segments, making the problem harder both because several transitions must be localized and because the number of changes is no longer fixed in advance. For the entropy and Frobenius methods, change-points are selected through a penalty-based stopping rule in \texttt{ruptures}' piecewise linear method \cite{truong_selective_2020}, using a small grid of positive penalties. For the LAD baseline, in the absence of an appropriate optimization scheme, we provide the additional information of the true number of change points and select the corresponding top anomalous points. Taken together, these three benchmarks probe complementary aspects of snapshot-based graph change-point detection: \emph{CommunityBench} tests sensitivity to structural reconfiguration at fixed activity, \emph{ActivityBench} tests sensitivity to pure activity changes under fixed structure, and \emph{MultiBench} tests robustness when several changes must be detected and their number is not prescribed.

\begin{figure}[htbp]
  \includegraphics[width=0.95\linewidth]{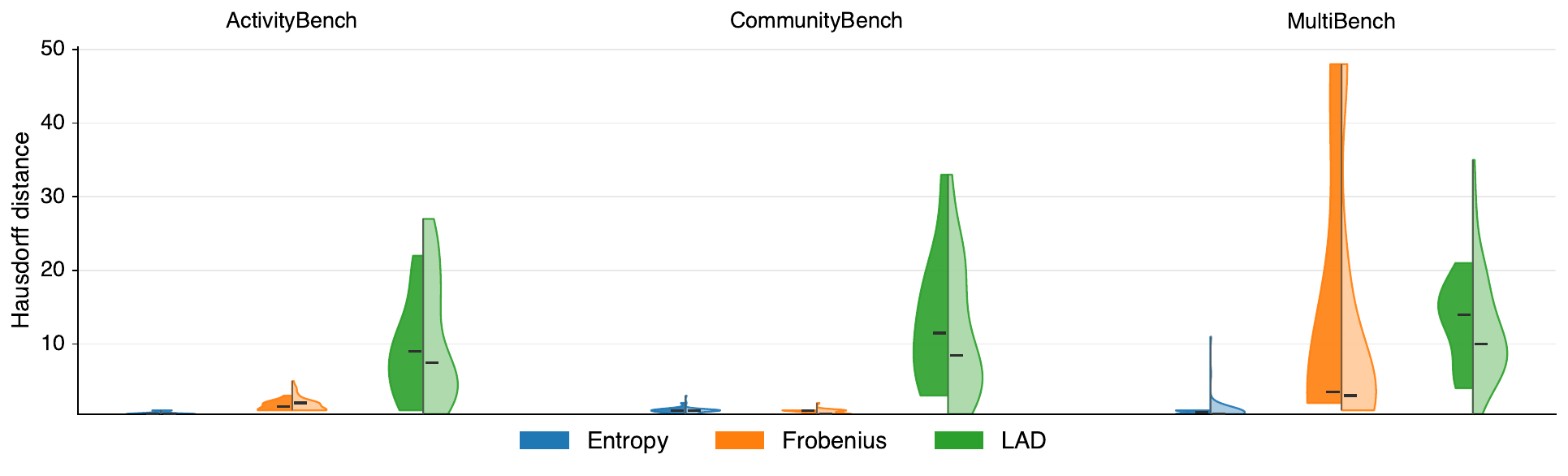}
  \caption{\label{fig_snapshot_experiments_train_test_split_violins}
  Per-sample Hausdorff distances for the snapshot-network benchmarks,
  comparing the entropy method with the Frobenius-distance and LAD baselines.
  For each method and each experiment, the left half of the violin (darker shade) shows the training-set distribution, while the right half (lighter shade) shows the test-set distribution. Lower values indicate more accurate localization of the ground-truth change points; the short horizontal segment within each half-violin marks the median, which is reported in \cref{tab:snapshot-hausdorff-comparison}.}
\end{figure}

Figure~\ref{fig_snapshot_experiments_train_test_split_violins} summarizes the results through the distribution of per-sample Hausdorff errors; their median values are collected in \cref{tab:snapshot-hausdorff-comparison}.In the \emph{CommunityBench} experiment, our entropy method and the Frobenius baseline are both competitive. The fact that the latter achieves the slightly smaller localization errors on both the training and test sets, indicates that direct snapshot differences already provide a strong signal when the transition is mainly structural. In contrast, on the \emph{ActivityBench} benchmark, our entropy method yields substantially smaller errors and more stable generalization from training to test, whereas LAD remains substantially less accurate. The same tendency persists, and becomes even more pronounced, in the \emph{MultiBench} setting: when several activity changes must be detected, the entropy signal remains the most robust of the three approaches, whereas both baselines degrade markedly, especially LAD, even while being informed of the true number of change points. Overall, these experiments suggest that the entropy-based construction is the most reliable method across heterogeneous snapshot scenarios, even though the simpler Frobenius comparison remains competitive on isolated single-change structural transitions. In \cref{fig:bkps_multibkps_block2activities_snapshots_first_sample_three_methods}, we display one representative example of the signals of the three methods for the \emph{MultiBench} training set.

\begin{table}[ht]
\centering
\begin{tabular}{llccc}
\toprule
 & & \multicolumn{3}{c}{Median Hausdorff Distance} \\
\cmidrule{3-5}
Split & Method & ActivityBench & CommunityBench & MultiBench \\
\midrule
Training & Entropy & 0.000 & 1.000  & 0.500  \\
 & Frobenius & 1.500  & 1.000  & 3.500  \\
 & LAD & 9.000  & 11.500  & 14.000  \\
\midrule
Test & Entropy & 0.000 & 1.000  & 0.000 \\
 & Frobenius & 2.000  & 0.000  & 3.000 \\
 & LAD & 7.500  & 8.500  & 10.000 \\
\bottomrule
\end{tabular}
\caption{Median localization error for the snapshot benchmarks.}
\label{tab:snapshot-hausdorff-comparison}
\end{table}

\section{\label{sec:Applications} Application to Community Detection}
Identifying change points has multiple downstream applications.  Here, we discuss their relevance for community detection. The performance of existing dynamic community detection methods increases when applied to structurally stable time intervals. Specifically, the precision benefits from avoiding considering together links that characterize different events, and the recall benefits from focusing on time windows where community structures are well-defined, therefore less noisy.
These benefits are not limited to clustering algorithms for static graphs on aggregated snapshots, but also to methods such as flow stability \cite{bovet_flow_2022}, which can suffer from false negatives due to interference from another community.

We therefore suggest performing a graph change-point detection step using our entropy-based approach before applying multiple dynamic community detection algorithms to the stable time intervals.
We now apply this methodology to the French primary school dataset \cite{stehle_high-resolution_2011}, a well-known real-world dataset. For dynamic community detection, we rely on the flow stability method \cite{bovet_flow_2022}. This method is thematically connected, as it also uses inhomogeneous transition kernels \cref{eq:inhomogeneous_kernel}. Moreover, they share the same computationally expensive step of computing inter-transition matrices \cref{eq:inter_T}, so one can compute and store them during the change-point detection phase and reuse them for community detection later, making the two methods complementary. Furthermore, the primary school network was previously investigated by Bovet et al. \cite{bovet_flow_2022} using flow stability, enabling direct assessment of the improvements in quality and explainability achieved by combining the method with the change-point detection step.

The French primary school dataset was generated by recording face-to-face contacts between 232 children and 10 teachers over two days using radio-frequency identification devices, worn on participants' chests, at 20-s resolution. Students are distributed across five grades with two classes per grade. Each class is assigned a room and a teacher. This separation usually restricts contact; however, during morning, lunch, and afternoon breaks, children mix in the playground or in the canteen. These common spaces do not have sufficient capacity to accommodate all students simultaneously; therefore, only two or three classes are on break at the same time, and lunches are taken in two consecutive turns \cite{bovet_flow_2022}.

\begin{figure}[htbp]
  \includegraphics[width=0.95\linewidth]{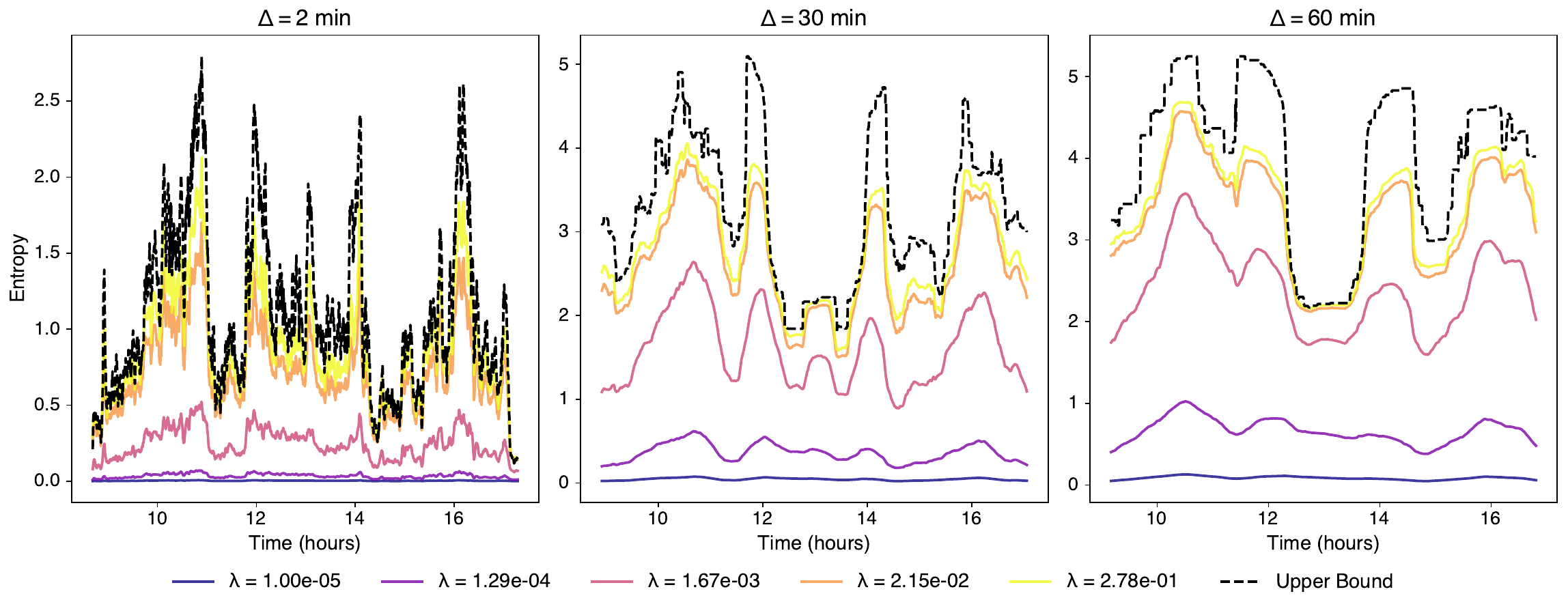}
  \caption{\label{fig:primary_school_windows} Local conditional entropy of diffusion for day 1 of the French primary school dataset. Every plot corresponds to a different rolling window, and every color to a different diffusion rate. Faster rates and smaller windows are noisier but more precise. Focusing on a single plot, faster rates appear to converge to an asymptotic curve that encodes the sizes of the connected components.}
\end{figure}

First, we compute local conditional entropy curves for various time windows and diffusion rates. Without any prior information, one must decide which context windows to investigate and which diffusion rates to use, so that the diffusion process can explore different scales within those windows and thereby modulate the signal-to-noise ratio. In this case, we are interested in short, intermediate, and long-lasting communities, so we scan entropies over windows of 2, 30, and 60 minutes and display the results in \cref{fig:primary_school_windows}. By inspecting the first panel, we observe entropy peaks at 10:30, 12:00, 14:00, and 16:00 that persist even when considering a longer context window $\Delta$. The ones at 12:00 and 14:00 correspond to the start and end of the lunch break and are particularly sharp. On the other hand, those around 10:30 and 16:00 correspond to recesses, and their shapes are more jagged, reflecting a rapid, complex evolution of the communities at those times. For fast diffusion rates, the value of the peaks is influenced by the size of the connected components of the underlying footprints $G_{[t-\frac{\Delta}{2},\,t+\frac{\Delta}{2}]}$ around those peaks, and the level of (a)symmetry of temporal paths. This allows us to compare the importance of various moments of the day in terms of structural activity. For instance, in this example, the morning recess exhibits richer network evolution than the afternoon one and undergoes more complex dynamics than at the start of the lunch break, the other major event of the day. Additionally, as argued before, signals generated with faster diffusion rates and smaller windows are noisier but more precise. For instance, when considering $\Delta = 60$ min signals, it is no longer clear exactly when lunch starts, and the lesser event at 13:00 is no longer detected, as it lasts less than the context window.
Here, the signal obtained with a 30 min window provides a detailed trace of the changes happening in this primary school, including some changes that are more gradual. For example, the increase starting around 9:45, culminating around 10:30, and then decreasing until around 11:15 captures the complex movements of classes from their classrooms to the playground for recess. Some classes begin moving to the playground at 9:45; among them, some return to their classrooms at 10:30, while others join the playground at the same time \cite{stehle_high-resolution_2011}, marking the time when many students cross paths and mixing is highest. This is also when the entropy signal reaches its maximum.
During the lunch break, many students leave the school, while others remain. This occurs around 11:00, and they return around 14:15, corresponding to two periods when many students mix and cross paths and are captured by a sudden increase in local entropy. The students remaining in school from two groups, alternating between the playground and the cafeteria, from 11:30 to 13:00, and then from 13:00 to 14:45. The increase in signal we observe around 13:00 again captures the moment when students cross paths between the playground and the cafeteria.

\begin{figure}[htbp]
  \includegraphics[width=0.95\linewidth]{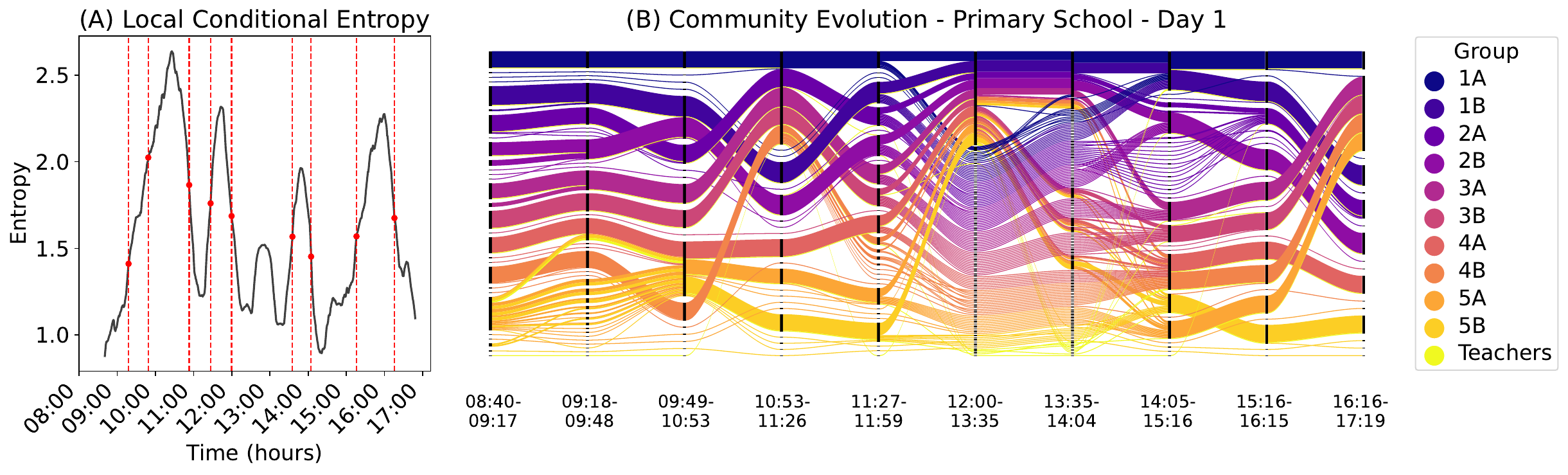}  \caption{\label{fig:entropy_inf_community_wide_sankey} Community detection of day 1 of the primary school dataset, on split time interval determined based on the local conditional entropy curve. (A) We rely on the local conditional entropy curve $H(t, 1.67 \times 10^{-3}, 30 \text{min})$ and detect change points with the \texttt{ruptures} algorithm with its piecewise linear model and a penalty term equal to 8. The detected change points split the day into the start of the day, the morning recess plus the return to classes, the lunch break, the early afternoon classes, and finally what comes after the afternoon recess. (B) Sankey diagram of communities detected with the flow stability method \cite{bovet_flow_2022} (1 scale for each sub-interval). Various interesting community dynamics emerge, like the morning and afternoon recesses, plus the lunch break.}
\end{figure}

Secondly, we focus on a single window length to match our desired time resolution for detectable communities, and pick one specific entropy curve and segment it as shown in \cref{fig:entropy_inf_community_wide_sankey}(A). In this example, we consider the 30-minute context window and select the signal associated with the diffusion rate of $1.67 \times 10^{-3}$ s$^{-1}$; we then segment it using the piecewise linear model from \texttt{ruptures} with a penalty term equal to 8. We identify the 9 time points 09:17, 09:48, 10:53, 11:27, 11:59, 13:35, 14:05, 15:16, and 16:15 as major change points. 

We then perform (multiscale) community detection on each sub-interval with flow stability and select a meaningful clustering using the average normalized variation of information heuristic, as shown in \cref{fig:primaryschool_interval_nvi_selection}.

We plot the evolution of communities from one sub-interval to the next one with a Sankey diagram \cref{fig:entropy_inf_community_wide_sankey}(B), as previously done in \cite{matias_statistical_2017}. This offers a clear picture of which communities exist, and when they appear and disappear. Further analysis of the relationships among detected communities is possible using topological data analysis tools, such as persistent homology \cite{schindler_analysing_2025,schindler2025mcbif}.

Beyond improving the interpretability of the clustering results, this approach makes it easier to verify their accuracy: after observing interesting patterns in the diagram, we can investigate the data and interpret them. As noted in the dataset description, children primarily interact with classmates. This is evident in the early-morning partition, with the notable exception of classes 5A and 5B, which are clustered together, which is consistent with other analyses of this network \cite{stehle_high-resolution_2011,bovet_flow_2022}. The detected clusters within the interval 09:49-10:53 capture the dynamics of the recess period and align with the order of arrival at the playground. The effects of the lunch break on community structure can be observed in two time intervals. From 12:00 to 13:35, the community structure consists of a single large cluster containing people from different classes, and mostly singletons, i.e., those who have lunch at school and those who go home. During the 13:35-14:04 interval, the large cluster splits, suggesting that, at least in the second part of the lunch break, students staying at school are somewhat fragmented. Finally, later intervals capture the return to class and the afternoon recess.
This example illustrates how combining the local entropy signal with change-point and community detection enables one to obtain a clear picture from a complex temporal network.

\section{\label{sec:DiscussionConclusions} Discussion and Conclusions}
Diffusion-based methods on graphs have played a central role in network science, graph signal processing and graph machine learning \cite{rosvall_maps_2008,delvenne_stability_2010,lambiotte_random_2014,bovet_flow_2022, thanou_learning_2017, xu_graph_2019}. In parallel, there has been a sustained interest in the entropic properties of diffusion and Markovian processes on networks. Early contributions include Refs. \cite{latora_kolmogorov-sinai_1999,gomez-gardenes_entropy_2008}, while more recent works \cite{villegas_laplacian_2022,villegas_laplacian_2023,nartallo-kaluarachchi_broken_2024} reflect renewed attention to the interplay between network structure, diffusion, and information-theoretic observables.

In this context, we introduced the study of conditional entropy for heat diffusion on static graphs \cite{koovely_evolution_2025}, together with a thermodynamic interpretation mirroring classical heat diffusion in Euclidean spaces. The present work extends this framework to dynamic networks. At the theoretical level, we generalized diffusion conditional entropy to link streams and studied its main properties. In particular, we showed that temporal conditional entropy remains monotone in time, identified natural upper bounds based on footprint connectivity, and clarified how its behavior differs from the static case, notably through the loss of monotonicity with respect to the diffusion rate in full generality. We then introduced a local version of conditional entropy, designed to probe diffusion over finite temporal windows and therefore better suited to capturing structural changes. This local viewpoint is the key methodological contribution of the paper. It transforms a temporal network into a low-dimensional signal that remains sensitive both to variations in activity and to changes in mesoscopic organization. From this perspective, the proposed method belongs to the broad family of signal-extraction approaches to graph change-point detection, but differs from existing methods in that the signal is derived from the way diffusion pathways evolve over time. This makes the approach particularly natural for link streams, where the temporal ordering and duration of contacts matter explicitly.

Our experiments support this interpretation on several levels. First, on synthetic continuous-time benchmarks, the local conditional entropy yields informative one-dimensional signals for both activity-driven and structure-driven changes. Second, on snapshot-network benchmarks, the resulting change-point detection pipeline performs competitively against two representative baselines, namely a direct Frobenius-distance method and the spectral LAD approach. The picture that emerges is nuanced: direct matrix distances remain highly effective in some simple single-change structural settings, whereas the entropy-based signal appears more robust across heterogeneous scenarios, especially when activity changes or multiple breakpoints are involved. Third, on the French primary school dataset, we showed that the detected change-points can be used to partition the temporal network into more stable intervals, thereby improving the interpretability of subsequent community detection.

More generally, this work illustrates how ideas from statistical physics can inform data analysis and machine learning on structured systems. Classic examples of this interaction include Hopfield networks \cite{hopfield_neural_1982,hopfield_nobel_2025} and restricted Boltzmann machines \cite{ackley_learning_1985,hinton_nobel_2025}. In our case, the central object is a diffusion process, but the final output is a signal that can be processed with standard segmentation tools. This bridge between diffusion physics and signal processing is, we believe, one of the framework's main conceptual strengths.

Several research directions follow from this work. First, although the method was tested against strong and relevant baselines in the snapshot setting, a broader empirical comparison would still be valuable. Such a comparison would be a natural next step. This is particularly important because the present approach applies natively to link streams, a setting in which many graph change-point detection methods are not directly usable.

Second, we focused on offline change-point detection. Extending the methodology to the online setting would be highly desirable, especially for applications in monitoring and control, where changes must be identified as data arrive.

Third, the experiments treated changes in activity and changes in community structure mostly as separate benchmark families. A natural extension would be to detect multiple types of changes simultaneously. One possible route is to replace the scalar entropy signal by a multivariate one obtained from several pairs $(\lambda,\Delta)$, so that different coordinates may become sensitive to different types or magnitudes of change.

Fourth, in addition to snapshot networks and link streams with nonzero link durations, one may also wish to study link streams with instantaneous interactions. In that setting, diffusion is not naturally adapted to the data, except perhaps in the limiting regime $\lambda\to\infty$, whose existence is currently only suggested empirically in our framework and has not been established rigorously. A possible approximation would be to assign each interaction a small artificial duration, shorter than the shortest inter-event time in the data, and then compute the entropy for a sufficiently large diffusion rate. Whether this approximation yields a stable and meaningful signal remains an open question.

Finally, it would also be worthwhile to investigate alternative diffusion operators, such as random-walk diffusion, whose linearized form may be easier to compute and could provide a more scalable basis for large-scale applications.

\paragraph*{Code availability.}
The code used to reproduce the figures, simulations, and experiments presented in this article is available at \url{https://github.com/samuelkoovely/ChangePointDetection}. It includes the implementation of the entropy-based change-point detection pipeline, the synthetic-data generation procedures, and the scripts used for the community-detection application. For the implementation of the Frobenius baseline, we adapted the code available at \url{https://github.com/dsulem/DyNNet}; for the LAD baseline we adapted the one from \url{https://github.com/shenyangHuang/LAD}. For segmentation we relied on the \texttt{ruptures} python library; the code is available at \url{https://github.com/deepcharles/ruptures}.

\section*{Acknowledgments}
We would like to acknowledge C. Truong for suggestions on designing the experimental setup and for explanations of \texttt{ruptures}. We also thank Y. Asgari and D. Quelle for suggesting improvements to the code implementation of the local conditional entropy.

\appendix

\section{Von Neumann Entropy}  \label{appendix:vNS}
The notion of Von Neumann entropy extends the notion of Gibbs entropy from statistical mechanics to quantum statistical mechanics. It can be considered the counterpart of Shannon entropy in quantum information theory. Later works have adapted other notions from information theory, e.g., conditional entropy, to the quantum world \cite{cerf_negative_1997, cerf_quantum_1999}. The spectral entropy described in \cite{de_domenico_spectral_2016, ghavasieh_statistical_2020} is an adaptation of von Neumann entropy to graphs and has been used to define a renormalization group for complex networks \cite{villegas_laplacian_2022, villegas_laplacian_2023} and for community detection \cite{villegas_multi-scale_2025}.

In this appendix, we describe this notion of spectral entropy and why it doesn't extend naturally to temporal networks.

A density matrix $\matr{R}$ is a Hermitian and positive semidefinite matrix with a trace equal to one. The von Neumann entropy of a density matrix $\matr{R}$ is defined as:
\begin{equation}
    S(\matr{R}) \coloneqq - \Tr(\matr{R} \log \matr{R}).
\end{equation}

Let $\{ \lambda_1, ..., \lambda_n \}$ be the set of eigenvalues of $\matr{R}$. One can show that
\begin{equation}
    S(\matr{R})= - \sum_{i=1}^N \lambda_i \log (\lambda_i).
\end{equation}

The quantity is well-defined because the eigenvalues are all non-negative real numbers. Moreover, the trace of a density matrix is by definition equal to one, so the set of eigenvalues constitutes a discrete probability distribution. We can therefore interpret the von Neumann entropy of a density matrix as the Shannon entropy of its eigenvalues' probability distribution.

Now, one can consider a static network as complex system whose structural information is encoded in the heat kernel $e^{- \Lapl t}$. We can then define the graph von Neumann entropy based on the density matrix
\begin{equation}
    \matr{R} (\tau) \coloneqq \frac{e^{-\tau L}}{Z},
\end{equation}
 with trace-normalizing constant $Z \coloneqq \sum_{i=1}^N e^{- \lambda_i}$, and compute it as the Shannon entropy of its eigenvalues.

In order for this construction to extend to dynamic networks, one would need the inhomogeneous heat transition kernel $\T(0, t)$ (\cref{eq:inhomogeneous_kernel}) to have a real non-negative spectrum as well. This is true in specific cases (e.g., if the kernel is obtained by multiplying two inter-transition matrices or any arbitrary number of commuting ones), but not in general. The temporal cycle from \cref{fig:Counterexample} considered previously is a counterexample: the transition kernel of this small system has eigenvalues that are not in $\mathbb{R}^+$. The spectral entropy, therefore, does not generalize naturally to temporal networks. We note, however, that there are recent works in quantum mechanics that propose generalizations of the von Neumann entanglement entropy for non-Hermitian transition matrices \cite{parzygnatSVDEntanglementEntropy2023,caputaMusingsSVDPseudo2024,chen_entropy_2025}.

\section{Fully Ordered Temporal Path}  \label{appendix:path_graph}
In the fully ordered temporal path $L_{N,0}$, the asymptotic value in the diffusion rate of its heat kernel $T^\infty_{N,0}$, from time 0 to $N-1$ , is obtained by composing the averaging effects associated with the links $\{1,2\},\{2,3\},\dots,\{N-1,N\}$ in this specific temporal order. The resulting matrix is upper triangular up to the
first two identical rows, and is given by
\begin{equation}
\label{eq:asymptotic_ordered_temporal_path_matrix}
T^\infty_{N,0}
=
\begin{pmatrix}
\frac12 & \frac14 & \frac18 & \cdots & \frac{1}{2^{N-1}} & \frac{1}{2^{N-1}} \\
\frac12 & \frac14 & \frac18 & \cdots & \frac{1}{2^{N-1}} & \frac{1}{2^{N-1}} \\
0 & \frac12 & \frac14 & \cdots & \frac{1}{2^{N-2}} & \frac{1}{2^{N-2}} \\
0 & 0 & \frac12 & \cdots & \frac{1}{2^{N-3}} & \frac{1}{2^{N-3}} \\
\vdots & \vdots & \vdots & \ddots & \vdots & \vdots \\
0 & 0 & 0 & \cdots & \frac12 & \frac12
\end{pmatrix}.
\end{equation}
Equivalently, the first two rows are equal, while for every row index
$r \in \{2,\dots,N\}$,
\begin{equation}
\label{eq:asymptotic_ordered_temporal_path_entries}
T^\infty_{N,0}(r,j)
=
\begin{cases}
0, & j < r-1, \\[0.4em]
2^{-(j-r+2)}, & r-1 \leq j \leq N-1, \\[0.4em]
2^{-(N-r+1)}, & j=N,
\end{cases}
\end{equation}
and $T^\infty_{N,0}(1,\cdot) = T^\infty_{N,0}(2,\cdot)$.

Hence the nonzero entries of row $r$, for $r \geq 2$, are
$\frac12,\frac14,\dots,
    \frac{1}{2^{N-r+1}},
    \frac{1}{2^{N-r+1}};
$ where the repetition of the final term comes from the last averaging operation: the remaining mass at vertex $N-1$ is split uniformly between $N-1$ and $N$. We can then compute the entropy of each row explicitly. Let
\begin{equation*}
    H_r \coloneqq - \sum_{j=1}^N T^\infty_{N,0}(r,j) \log T^\infty_{N,0}(r,j)
\end{equation*}
be the entropy of the $r$-th row. Since all
nonzero entries are powers of $1/2$, an entry of the form $2^{-\ell}$
contributes
\begin{equation*}
    -2^{-\ell}\log(2^{-\ell})
    =
    \frac{\ell \log (2)}{2^\ell}.
\end{equation*}
Therefore, for $r\geq 2$, setting
\begin{equation*}
    m=N-r+1,
\end{equation*}
we obtain
\begin{equation*}
    H_r = \log(2) \left(\sum_{\ell=1}^{m} \frac{\ell}{2^\ell} + \frac{m}{2^m} \right).
\end{equation*}
Using
\begin{equation*}
    \sum_{\ell=1}^{m} \frac{\ell}{2^\ell} = 2-\frac{m+2}{2^m},
\end{equation*}
this simplifies to
\begin{equation}
\label{eq:asymptotic_ordered_temporal_path_row_entropy}
    H_r = \log(2) \left(2-\frac{2}{2^m} \right)
    = \log(2) \left(2-2^{r-N} \right), \qquad r\geq 2.
\end{equation}
Since the first two rows of $T^\infty_{N,0}$ are identical, we also have
\begin{equation*}
    H_1= H_2 = \log(2) \left(2-2^{2-N} \right).
\end{equation*}

Finally, if the initial distribution is uniform on $V_N$, the conditional
entropy of the limiting transition matrix is the average of the row
entropies: $ H(T^\infty_{N,0}\mid Z(0)) = \frac{1}{N} \sum_{r=1}^{N} H_r$.
So, substituting \cref{eq:asymptotic_ordered_temporal_path_row_entropy} gives
\begin{align}
    H(T^\infty_{N,0}\mid Z(0))
    &= \frac{\log(2)}{N} \left[2\left(2-2^{2-N}\right) + \sum_{r=3}^{N} \left(2 - 2^{r-N}\right) \right] \notag \\
    &= \left(2-\frac{2}{N}\right)\log(2).
    \label{eq:asymptotic_ordered_temporal_path_global_entropy}
\end{align}

\begin{figure}[htbp]
  \includegraphics[width=0.95\linewidth]{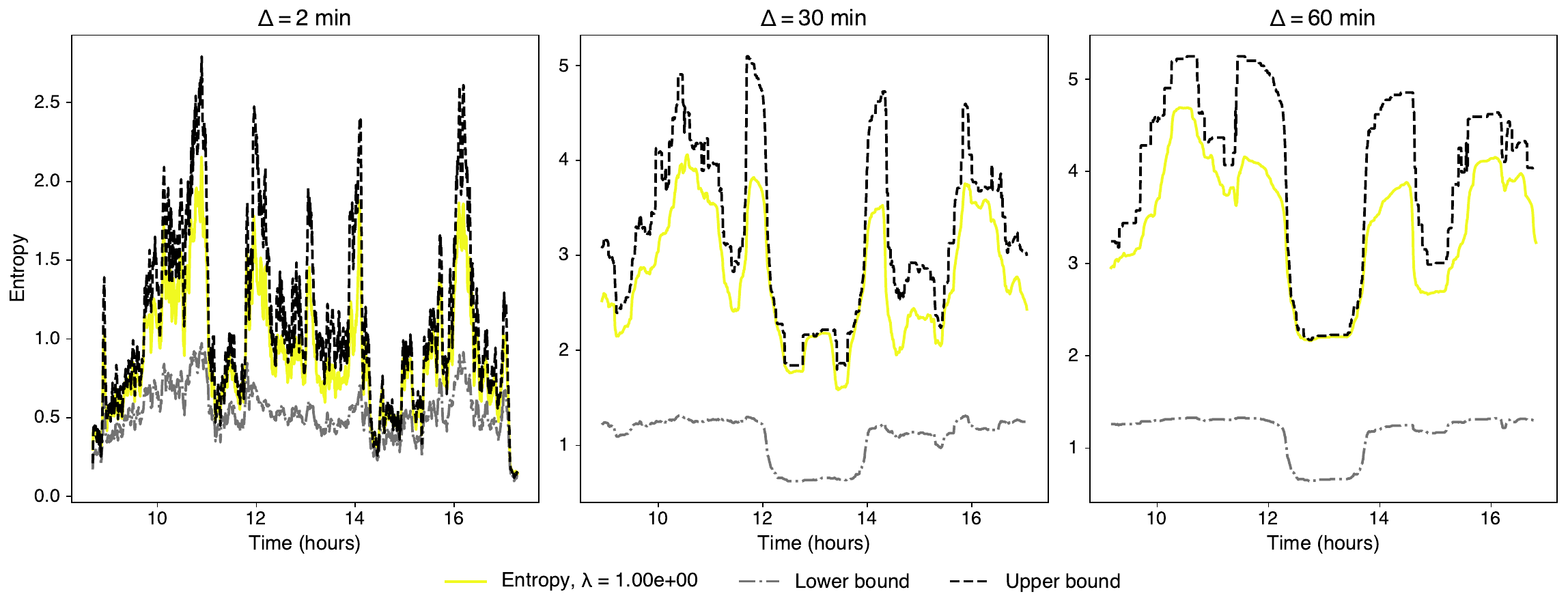}
  \caption{\label{fig:primaryschool_fastest_bounds} Primary-school day-1 entropy signal for the fastest diffusion rate,
$\lambda = 1$, shown together upper and lower bounds for $\Delta = 2$, $30$, and $60$ minutes. The upper bound is $ \sum_{k=1}^\ell \frac{|V_k|}{N}\log(|V_k|)$, whereas the lower bound is $ \sum_{k=1}^\ell \frac{|V_k|}{N}\left(2-\frac{2}{|V_k|}\right)\log(2)$, where $k$ ranges over all the connected components of the graph aggregated over each time window. For the shortest window, the components are smaller and the two bounds are therefore much tighter. For the longer windows, larger connected components make the upper bound increase while the lower bound remains comparatively low, producing a wider admissible range.}
\end{figure}

Thus, in the fully ordered temporal path $L_{N,0}$, the limiting final
conditional entropy is
\begin{equation*}
    H(T^\infty_{N,0}\mid Z(0)) = \left(2-\frac{2}{N}\right)\log(2).
\end{equation*}
Based on this value, we propose a lower bound for the asymptotic local entropy curve by taking \cref{eq:lower_bound} over the connected components of the local footprint $G_{[t-\frac{\Delta}{2},\,t+\frac{\Delta}{2}]}$. We capture the quality of this bound in \cref{fig:primaryschool_fastest_bounds}, where we display it together with the entropy curve associated to the fastest diffusion displayed in \cref{fig:primary_school_windows} together with its upper bound estimate. Entropy values close to the upper bound indicate that asymptotic transition matrices of heat diffusion over the static footprints are closer to being symmetric, while values close to the lower bound indicate stronger asymmetry.

\section{Additional Figures}
Here, we put additional figures that complement the ones in the main text.
\begin{figure}[htbp]   \includegraphics[width=0.95\linewidth]{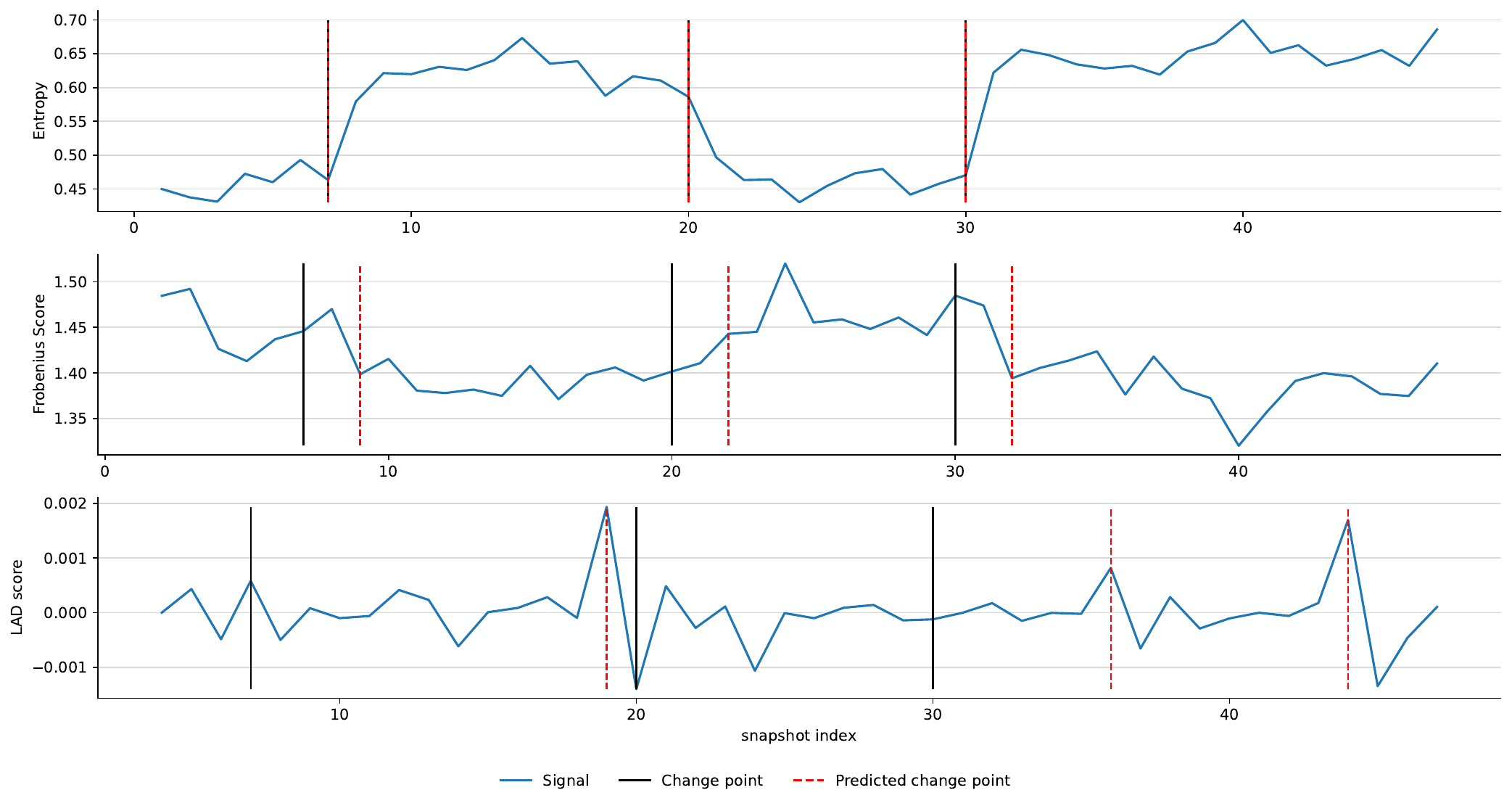}
    \caption{\label{fig:bkps_multibkps_block2activities_snapshots_first_sample_three_methods} Comparison of the signals' quality of the three different tested methods, for the first sample of the \emph{Benchmark3} training set. For each method, we display the signal and detected change points according to the optimal parameterization. In this specific case, there are three true change points, and the entropy signal allows to recover them with perfect precision. The Frobenius baseline is also competitive, but the signal is less interpretable to the human eye. The LAD baseline is not able to recover the three change points successfully, despite being informed of their actual number.}
\end{figure}

\begin{figure}[htbp]   \includegraphics[width=0.75\linewidth]{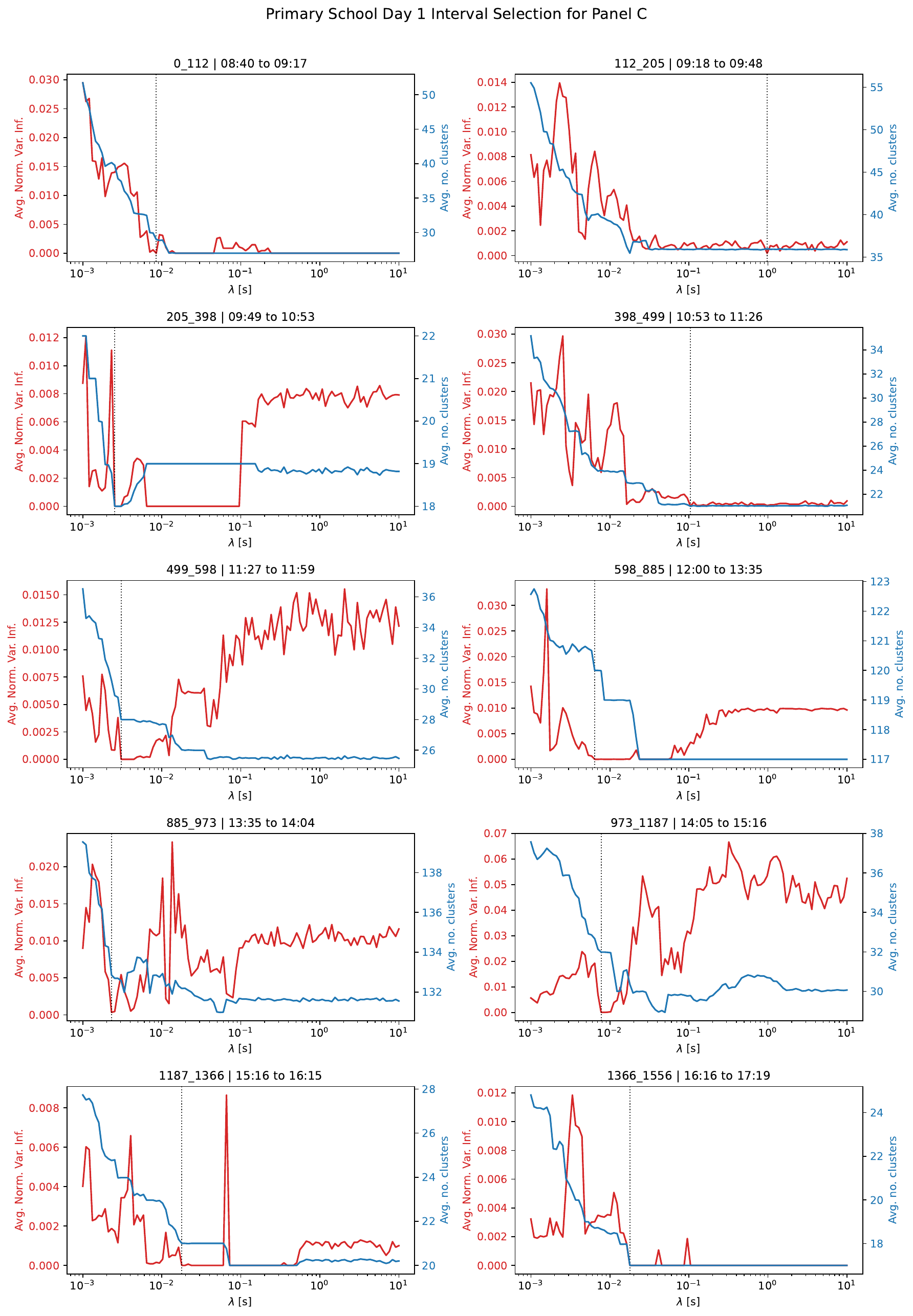}
\centering
    \caption{\label{fig:primaryschool_interval_nvi_selection} Multiscale-community detection on a sub-interval with flow stability \cite{bovet_flow_2022}. The selected scales, the minima of the average normalized variation of information curve, are highlighted with black dotted lines. }
\end{figure}

\bibliographystyle{ieeetr}
\bibliography{bibliography}

\end{document}